\documentclass[runningheads]{llncs}
\pdfoutput=1
\usepackage[utf8]{inputenc}
\usepackage{graphicx}
\usepackage{amsmath}
\usepackage{amssymb}
\usepackage{tikz}
\usepackage{pgfplots}
\usepackage[vlined,linesnumbered]{algorithm2e}
\usepackage{hyperref}
\usepackage{multirow}
\usepackage{wrapfig}

\title{Geometric Total Variation for Image Vectorization, Zooming and
  Pixel Art Depixelizing \thanks{This work has been partly funded by 
\textsc{CoMeDiC} ANR-15-CE40-0006 research grant.}}

\titlerunning{Geometric Total Variation}
\author{Bertrand Kerautret\inst{1}
  \and
  Jacques-Olivier Lachaud\inst{2}
}
\authorrunning{B. Kerautret and J.-O. Lachaud}

\institute{Univ. Lyon 2, LIRIS, CNRS, France\\
  \email{bertrand.kerautret@liris.cnrs.fr}\\           
  \and
  Univ. Grenoble Alpes, Univ. Savoie Mont Blanc\\
  CNRS, LAMA, 73000 Chambéry, France\\
  \email{jacques-olivier.lachaud@univ-smb.fr}           
}

\newcommand{\etal}{{\it et~al.}}
\newcommand{\R}{\ensuremath{\mathbb{R}}}

\newcommand{\D}{\ensuremath{\mathrm{d}}}

\newcommand{\Grad}{\ensuremath{\nabla}}
\newcommand{\DGrad}{\ensuremath{\boldsymbol{\nabla}}}
\newcommand{\Interval}[2]{\ensuremath{\lbrack #1, #2 \rbrack}}
\newcommand{\TOmega}{{\ensuremath{\mathcal{T}(\Omega)}}}
\newcommand{\NormK}[1]{\ensuremath{\left\| #1 \right\|_K}}
\newcommand{\TV}{\ensuremath{\mathrm{TV}}}
\newcommand{\GTV}{\ensuremath{\mathrm{GTV}}}
\newcommand{\vp}{\ensuremath{\mathbf{p}}}
\newcommand{\vq}{\ensuremath{\mathbf{q}}}
\newcommand{\vr}{\ensuremath{\mathbf{r}}}
\newcommand{\vx}{\ensuremath{\mathbf{x}}}
\newcommand{\vb}{\ensuremath{\mathbf{b}}}
\newcommand{\vy}{\ensuremath{\mathbf{y}}}
\newcommand{\vz}{\ensuremath{\mathbf{z}}}
\newcommand{\vT}{\ensuremath{\mathbf{T}}}
\newcommand{\Eq}[1]{(\ref{#1})}
\newcommand{\RefFigure}[1]{Figure~\ref{#1}}
\newcommand{\RefSection}[1]{Section~\ref{#1}}
\newcommand{\RefAlgorithm}[1]{Algorithm~\ref{#1}}
\newcommand{\RefLine}[1]{line~\ref{#1}}

\SetKw{fct}{function}
\SetKw{ret}{return}
\SetKw{var}{var}

\newcommand{\Face}[1]{\ensuremath{\mathrm{face}(#1)}}
\newcommand{\FaceP}[1]{\ensuremath{\mathrm{face}'(#1)}}
\newcommand{\Injz}{\ensuremath{\iota_z}}
\newcommand{\Vor}[1]{\ensuremath{\mathrm{Vor}(#1)}}

\newcommand{\Paragraph}[1]{\noindent\emph{#1}}

\begin{document}
\maketitle

\begin{abstract}
  We propose an original method for vectorizing an image or zooming it
  at an arbitrary scale. The core of our method relies on the
  resolution of a geometric variational model and therefore offers
  theoretic guarantees. More precisely, it associates a total
  variation energy to every valid triangulation of the image
  pixels. Its minimization induces a triangulation that reflects image
  gradients. We then exploit this triangulation to precisely locate
  discontinuities, which can then simply be vectorized or zoomed.
  This new approach works on arbitrary images without any learning
  phase. It is particularly appealing for processing images with low
  quantization like pixel art and can be used for depixelizing such
  images. The method can be evaluated with an online demonstrator,
  where users can reproduce results presented here or upload their own
  images.

\end{abstract}

\section{Introduction}

There exist many methods for zooming into raster
images, and they can be divided into two groups. The first group
gathers ``image super-resolution'' or ``image interpolation''
techniques and tries to deduce colors on the zoomed image from nearby
colors in the original image. These methods compute in the raster
domain and their output is not scalable. The second group is composed
of ``image vectorization'' approaches and tries to build a higher
level representation of the image made of painted polygons and
splines. These representations can be zoomed, rotated or
transformed. Each approach has its own merits, but none perform well
both on cmaera pictures and on drawings or pixel art.

\Paragraph{Image interpolation/super resolution.}

These approaches construct an interpolated image with higher
resolution. A common trait is that they see the image as a continuous
function, such that we only see its sampling in the original
image. Their diversity lies generally in the choice of the space of
continuous functions. For instance linear/cubic interpolation just
fits linear/cubic functions, but much more complicated spaces have
been considered. For instance the method proposed by Getreuer
introduces the concept of \textit{contour stencils}
\cite{getreuer_contour_2011,getreuer_image_2011bis} to accurately
estimate the edge orientation.  Roussos and Maragos proposed another
strategy using a tensor driven diffusion
\cite{roussos_vector-valued_2007} (see also
\cite{getreuer_roussos-maragos_2011}).

Due to the increasing popularity of convolutional neural network (CNN),
several authors propose super resolution networks to obtain super
resolution image. For instance Shi {\etal} define a CNN architecture
exploiting feature maps computed in low resolution
space \cite{shi_real-time_2016}. After learning, this strategy reduces
the computational complexity for a fixed super-resolution and gives
interesting results on photo pictures. Of course, all these approaches
do not provide any vectorial representation and give low quality
results on images with low quantization or pixel art images. Another
common feature is their tendency to ``invent'' false oscillating contours,
like in Fourier super-resolution.

\Paragraph{Pixel art super-resolution.} A subgroup of these methods is
dedicated to the interpolation of pixel art images, like the classic
$HqX$ magnification filter \cite{Stepin2003} where $X$ represents the
scale factor. We may quote a more advanced approach which proposes as
well a full vectorization of pixel art
image \cite{kopf_depixelizing_2011}. Even if the resulting
reconstruction for tiny images is nice looking, the proposed approach
is designed specifically for hand-crafted pixel art image and is not
adaptable to photo pictures.

\Paragraph{Image vectorization.}
Recovering a vector-based
representation from a bitmap image is a classical problem in the
field of digital geometry, computer graphics and pattern
recognition.
Various commercial design softwares propose such a feature,
like \textit{Indesign}{\texttrademark}, or Vector
Magic \cite{vectorMagic10} and their algorithms are naturally not
disclosed. Vectorization is also important for technical document
analysis and is often performed by decomposition into arcs and
straight line segments \cite{hilaire_robust_2006}. 
For binary images, many vectorization methods are available and may rely on dominant points detection
\cite{marji_polygonal_2004,Nguyen11}, relaxed straightness properties
\cite{bhowmick_fast_2007}, multi-scale analysis
\cite{feschet_multiscale_2010} or  digital curvature computation \cite{kerautret_curvature_2008,liu_unified_2008}.

Extension to grey level images is generally achieved by image decomposition into level sets, vectorization of each level, then fusion (e.g. \cite{KerautretNKV17}). Extension to color images is not straightforward.
Swaminarayan and Prasad proposed to use contour detection associated
to a Delaunay triangulation \cite{swaminarayan_rapid_2006} but despite
the original strategy, the digital geometry of the image is not
directly exploited like in above-mentionned approaches. Other methods
define the vectorization as a variational problem, like Lecot and Levy
who use the Mumford Shah piecewise smooth model and an intermediate
triangulation step to best approximate the image
\cite{lecot_ardeco:_2006}. Other comparable methods construct the
vectorization from topological preservation criteria
\cite{sun_image_2007}, from splines and adaptive Delaunay
triangulation \cite{demaret_image_2006} or from
Bézier-curves-patch-based representation
\cite{xia_patch-based_2009}. We may also cite the interactive method
proposed by Price and Barrett that let a user edit interactively the
reconstruction from a graph cut segmentation
\cite{price_object-based_2006}.

\Paragraph{Our contribution.}
We propose an original approach to this problem, which borrows and
mixes techniques from both groups of methods, thus making it very
versatile. We start with some notations that will be used thoughout
the paper.  Let $\Omega$ be some open subset of $\R^2$, here a
rectangular subset defining the image domain. Let $K$ be the image
range, a vector space that is generally $\R$ for monochromatic images
and $\R^3$ for color image, equipped with some norm $\NormK{ \cdot
}$. We assume in the following that gray-level and color components
lie in $\Interval{0}{1}$.

First our approach is related with the famous Total
Variation (\TV), a well known variational model for image denoising or
inpainting \cite{rudin_nonlinear_1992}.
Recall that if $f$ is a differentiable function in $\Omega$, its total variation
can be simply defined as:
\begin{equation}
  \TV(f) := \int_\Omega \NormK{ \Grad f (x) } \D x.
\end{equation}
In a more general setting, the total variation is generally defined by
duality. As noted by many authors \cite{Condat:2017-siims}, different
discretizations of $\TV$ are not equivalent. However they are all
defined locally with neighboring pixels, and this is why they are not
able to fully capture the structure of images.

In \RefSection{sec-geometric-tv}, we propose a {\em geometric total
variation}. Its optimization seeks the triangulation of lattice points
of $\Omega$ that minimizes a well-defined $\TV$ corresponding to a
piecewise linear function interpolating the image. The optimal
triangulation does not always connect neighboring pixels and align its
edges with image discontinuities. For instance digital straight
segments are recognized by the geometric total variation.

In \RefSection{sec-contour-reg}, we show how to construct a vector
representation from the obtained triangulation, by regularizing a kind
of graph dual to the triangulation. \RefSection{sec-2nd-order} shows a super-resolution algorithm that builds a zoomed raster image with smooth Bezier discontinuities from the vector representation. Finally, in \RefSection{sec-experiments} we compare our approach to other super-resolution and vectorization methods and discuss the results.

\section{Geometric Total Variation}
\label{sec-geometric-tv}

The main idea of our geometric total variation is to structure the
image domain into a triangulation, whose vertices are the pixel centers
located at integer coordinates. Furthermore, any triangle of this
triangulation should cover exactly three lattice points (its
vertices). The set of such triangulations of lattice points in
$\Omega$ is denoted by $\TOmega$.

Let $s$ be an image, {i.e.} a function from the set of lattice points
of $\Omega$ to $K$.
We define the {\em geometric total variation} of $s$ with respect to an
arbitrary triangulation $T$ of $\TOmega$ as
\begin{equation}
  \label{eq-gtv}
  \GTV(T, s) := \int_\Omega \NormK{ \Grad \Phi_{T,s}(x) } dx,
\end{equation}
where $\Phi_{T,s}(x)$ is the linear interpolation of $s$ in the
triangle of $T$ containing $x$. Note that points of $\Omega$ whose
gradient is not defined have null measure in the above integral (they
belong to triangle edges of $T$).

And the Geometric Total Variation (GTV) of $s$ is simply the smallest among
all possible triangulations:
\begin{equation}
  \GTV_{\TOmega} (s) := \min_{T \in \TOmega} \GTV(T,s).
\end{equation}

In other words, the GTV of a digital image $s$ is the smallest
continuous total variation among all natural triangulations sampling
$s$. Since $\TOmega$ will not change in the following, we will 
omit it as subscript and write simply $\GTV(s)$.

One should note that classical discrete TV models associated to a
digital image generally overestimate its total variation (in fact, the
perimeter of its level-sets due to the co-area formula)
\cite{Condat:2017-siims}. Classical discrete TV models $\TV(s)$ are
very close to $\GTV(T,s)$ when $T$ is any trivial Delaunay
triangulation of the lattice points. By finding more meaningful
triangulation, $\GTV(s)$ will often be much smaller than $\TV(s)$.

\Paragraph{Combinatorial expression of geometric total variation.}
Since $\Phi_{T,s}(x)$ is the linear interpolation of $s$ in the
triangle $\tau$ of $T$ containing $x$, it is clear that its gradient is
constant within the interior of $\tau$. If $\tau=\vp\vq\vr$ where $\vp, \vq,
\vr$ are the vertices of $\tau$, one computes easily that for any $x$ in
the interior of $\tau$, we have
\begin{align}\label{eq-grad-phi}
\Grad \Phi_{T,s}(x) & = s(\vp) (\vr -\vq)^\perp
  + s(\vq) (\vp -\vr)^\perp 
  + s(\vr) (\vq -\vp)^\perp. 
\end{align}
We thus define the discrete gradient of $s$ within a triangle $\vp\vq\vr$ as
\begin{align}
  \label{eq-def-dgrad}
\DGrad s (\vp\vq\vr)  := s(\vp) (\vr -\vq)^\perp
  + s(\vq) (\vp -\vr)^\perp 
  + s(\vr) (\vq -\vp)^\perp,
\end{align}
where $\vx^\perp$ is the vector $\vx$ rotated by $\frac{\pi}{2}$.
Now it is well known that any lattice triangle that has no integer
point in its interior has an area $\frac{1}{2}$ (just apply Pick's
theorem for instance). It follows that for any triangulation $T$ of
$\TOmega$, every triangle has the same area $\frac{1}{2}$. We obtain
the following expression for the $\GTV$:

\begin{align}
  \GTV(T, s) & = \int_\Omega \NormK{ \Grad \Phi_{T,s}(x) } dx \nonumber \\
  & = \sum_{\vp\vq\vr \in T} \int_{\vp\vq\vr} \NormK{ \Grad \Phi_{T,s}(x) } dx && \text{(integral is additive)} \nonumber \\
  & = \sum_{\vp\vq\vr \in T} \int_{\vp\vq\vr} \NormK{ \DGrad s (\vp\vq\vr) } dx && \text{(by \Eq{eq-grad-phi} and \Eq{eq-def-dgrad})} \nonumber \\
  & = \frac{1}{2}\sum_{\vp\vq\vr \in T} \NormK{ \DGrad s (\vp\vq\vr) } && \text{(since $\mathrm{Area}(\vp\vq\vr) = \frac{1}{2}$)}  \label{eq-gtv-expression}
\end{align}

\Paragraph{Minimization of \GTV.}
Since we have a local combinatorial method to compute the $\GTV$ of an
arbitrary triangulation, our approach to find the optimal $\GTV$ of some image
$s$ will be greedy, iterative and randomized. We will start from an arbitrary
triangulation (here the natural triangulation of the grid where
each square has been subdivided in two triangles) and we will flip two
triangles whenever it decreases the \GTV.

This approach is similar in spirit to the approach of Bobenko and
Springborn \cite{Bobenko:2007-dcg}, whose aim is to define discrete
minimal surfaces. However they minimize the Dirichlet energy of the
triangulation (i.e. squared norm of gradient) and they show that the
optimum is related to the intrinsic Delaunay triangulation of the
sampling points, which can be reached by local triangle flips. On the
contrary, our energy has not this nice convexity property and it is
easy to see for instance that minima are generally not unique. We have
also checked that there is a continuum of energy if we minimize the
$p$-norm $\NormK{\cdot}^p$ for $1 \le p \le 2$. For $p=2$ the optimum
is trivially the Delaunay triangulation of the lattice points. When
$p$ is decreased toward $1$, we observe that triangle edges are more
and more perpendicular to the discrete
gradient.

\RefFigure{fig-illustration-gtv} shows the $\GTV$ of several
triangulations in the simple case of a binary image representing an
edge contour with slope $\frac{2}{3}$. It shows that the smaller the
$\GTV$ the more align with the digital contour is the triangulation.

\begin{table}[t]
  \begin{tabular}{cccc}
    \rotatebox{90}{Triangulations}&
\pgfdeclareverticalshading{btw}{100bp}{color(0bp)=(black);color(25bp)=(black); color(75bp)=(white); color(100bp)=(white)}
\pgfdeclareverticalshading{btw_ldiag}{100bp}{color(0bp)=(black);color(20bp)=(black); color(45bp)=(white); color(100bp)=(white)}
\pgfdeclareverticalshading{btw_udiag}{100bp}{color(0bp)=(black);color(50bp)=(black); color(80bp)=(white); color(100bp)=(white)}
\begin{tikzpicture}
  \tikzstyle{blacktriangle}=[thick,red,fill=black]
  \tikzstyle{whitetriangle}=[thick,red,fill=white]
  \tikzstyle{shadedtriangle}=[thick,red,shade,shading=btw,shading angle=0]
  \tikzstyle{ldshadedtriangle}=[thick,red,shade,shading=btw_ldiag,shading angle=45]
  \tikzstyle{udshadedtriangle}=[thick,red,shade,shading=btw_udiag,shading angle=45]
  \draw[shadedtriangle] (0,1) -- (1,1) -- (0,0) -- cycle;
  \draw[shadedtriangle] (1,1) -- (0,0) -- (1,0) -- cycle;
  \draw[udshadedtriangle] (1,1) -- (2,1) -- (1,0) -- cycle;
  \draw[blacktriangle]  (2,1) -- (1,0) -- (2,0) -- cycle;
  \draw[blacktriangle]  (2,1) -- (3,1) -- (2,0) -- cycle;
  \draw[blacktriangle]  (3,1) -- (2,0) -- (3,0) -- cycle;
  \draw[whitetriangle]  (0,2) -- (1,2) -- (0,1) -- cycle;
  \draw[whitetriangle]  (1,2) -- (0,1) -- (1,1) -- cycle;
  \draw[whitetriangle]  (1,2) -- (2,2) -- (1,1) -- cycle;
  \draw[ldshadedtriangle] (2,2) -- (1,1) -- (2,1) -- cycle;
  \draw[udshadedtriangle] (2,2) -- (3,2) -- (2,1) -- cycle;
  \draw[blacktriangle]  (2,1) -- (3,1) -- (3,2) -- cycle;
  \tikzstyle{pixelnoir}=[black,fill=black]
  \tikzstyle{pixelblanc}=[black,fill=white]
  \draw[pixelnoir] (0,0) circle (0.1) (1,0) circle (0.1) (2,0) circle (0.1) (3,0) circle (0.1) (2,1) circle (0.1) (3,1) circle (0.1) (3,2) circle (0.1);
  \draw[pixelblanc] (0,1) circle (0.1) (1,1) circle (0.1) (0,2) circle (0.1) (1,2) circle (0.1) (2,2) circle (0.1);
\end{tikzpicture}







&
\pgfdeclareverticalshading{btw}{100bp}{color(0bp)=(black);color(25bp)=(black); color(75bp)=(white); color(100bp)=(white)}
\pgfdeclareverticalshading{btw_ldiag}{100bp}{color(0bp)=(black);color(20bp)=(black); color(45bp)=(white); color(100bp)=(white)}
\pgfdeclareverticalshading{btw_udiag}{100bp}{color(0bp)=(black);color(50bp)=(black); color(80bp)=(white); color(100bp)=(white)}
\pgfdeclareverticalshading{btw_uddiag}{100bp}{color(0bp)=(black);color(52bp)=(black); color(67bp)=(white); color(100bp)=(white)}
\begin{tikzpicture}
  \tikzstyle{blacktriangle}=[thick,red,fill=black]
  \tikzstyle{whitetriangle}=[thick,red,fill=white]
  \tikzstyle{shadedtriangle}=[thick,red,shade,shading=btw,shading angle=0]
  \tikzstyle{ldshadedtriangle}=[thick,red,shade,shading=btw_ldiag,shading angle=45]
  \tikzstyle{udshadedtriangle}=[thick,red,shade,shading=btw_udiag,shading angle=45]
  \tikzstyle{uddshadedtriangle}=[thick,red,shade,shading=btw_uddiag,shading angle=45]
  \draw[shadedtriangle] (0,1) -- (1,1) -- (0,0) -- cycle;

  \draw[uddshadedtriangle] (1,1) -- (2,1) -- (0,0) -- cycle;
  \draw[blacktriangle]  (0,0) -- (1,0) -- (2,1) -- cycle;

  \draw[blacktriangle]  (2,1) -- (1,0) -- (2,0) -- cycle;
  \draw[blacktriangle]  (2,1) -- (3,1) -- (2,0) -- cycle;
  \draw[blacktriangle]  (3,1) -- (2,0) -- (3,0) -- cycle;
  \draw[whitetriangle]  (0,2) -- (1,2) -- (0,1) -- cycle;
  \draw[whitetriangle]  (1,2) -- (0,1) -- (1,1) -- cycle;
  \draw[whitetriangle]  (1,2) -- (2,2) -- (1,1) -- cycle;
  \draw[ldshadedtriangle] (2,2) -- (1,1) -- (2,1) -- cycle;
  \draw[udshadedtriangle] (2,2) -- (3,2) -- (2,1) -- cycle;
  \draw[blacktriangle]  (2,1) -- (3,1) -- (3,2) -- cycle;
  \tikzstyle{pixelnoir}=[black,fill=black]
  \tikzstyle{pixelblanc}=[black,fill=white]
  \draw[pixelnoir] (0,0) circle (0.1) (1,0) circle (0.1) (2,0) circle (0.1) (3,0) circle (0.1) (2,1) circle (0.1) (3,1) circle (0.1) (3,2) circle (0.1);
  \draw[pixelblanc] (0,1) circle (0.1) (1,1) circle (0.1) (0,2) circle (0.1) (1,2) circle (0.1) (2,2) circle (0.1);
\end{tikzpicture}&
\pgfdeclareverticalshading{btw}{100bp}{color(0bp)=(black);color(25bp)=(black); color(75bp)=(white); color(100bp)=(white)}
\pgfdeclareverticalshading{btw_ldiag}{100bp}{color(0bp)=(black);color(20bp)=(black); color(45bp)=(white); color(100bp)=(white)}
\pgfdeclareverticalshading{btw_udiag}{100bp}{color(0bp)=(black);color(50bp)=(black); color(80bp)=(white); color(100bp)=(white)}
\pgfdeclareverticalshading{btw_uddiag}{100bp}{color(0bp)=(black);color(45bp)=(black); color(60bp)=(white); color(100bp)=(white)}
\pgfdeclareverticalshading{btw_udddiag}{100bp}{color(0bp)=(black);color(50bp)=(black); color(55bp)=(white); color(100bp)=(white)}
\begin{tikzpicture}
  \tikzstyle{blacktriangle}=[thick,red,fill=black]
  \tikzstyle{whitetriangle}=[thick,red,fill=white]
  \tikzstyle{shadedtriangle}=[thick,red,shade,shading=btw,shading angle=0]
  \tikzstyle{ldshadedtriangle}=[thick,red,shade,shading=btw_ldiag,shading angle=45]
  \tikzstyle{udshadedtriangle}=[thick,red,shade,shading=btw_udiag,shading angle=45]
  \tikzstyle{uddshadedtriangle}=[thick,red,shade,shading=btw_uddiag,shading angle=60]
  \tikzstyle{udddshadedtriangle}=[thick,red,shade,shading=btw_udddiag,shading angle=45]
  \draw[shadedtriangle] (0,1) -- (1,1) -- (0,0) -- cycle;

  \draw[udddshadedtriangle] (1,1) -- (3,2) -- (0,0) -- cycle;
  \draw[blacktriangle]  (0,0) -- (3,2) -- (2,1) -- cycle;

  \draw[blacktriangle]  (2,1) -- (1,0) -- (2,0) -- cycle;
  \draw[blacktriangle]  (2,1) -- (3,1) -- (2,0) -- cycle;
  \draw[blacktriangle]  (3,1) -- (2,0) -- (3,0) -- cycle;
  \draw[whitetriangle]  (0,2) -- (1,2) -- (0,1) -- cycle;
  \draw[whitetriangle]  (1,2) -- (0,1) -- (1,1) -- cycle;
  \draw[whitetriangle]  (1,2) -- (2,2) -- (1,1) -- cycle;

  \draw[uddshadedtriangle] (1,1) -- (2,2) -- (3,2) -- cycle;
  \draw[blacktriangle]  (0,0) -- (2,1) -- (1,0) -- cycle;
  
  \draw[blacktriangle]  (2,1) -- (3,1) -- (3,2) -- cycle;
  \tikzstyle{pixelnoir}=[black,fill=black]
  \tikzstyle{pixelblanc}=[black,fill=white]
  \draw[pixelnoir] (0,0) circle (0.1) (1,0) circle (0.1) (2,0) circle (0.1) (3,0) circle (0.1) (2,1) circle (0.1) (3,1) circle (0.1) (3,2) circle (0.1);
  \draw[pixelblanc] (0,1) circle (0.1) (1,1) circle (0.1) (0,2) circle (0.1) (1,2) circle (0.1) (2,2) circle (0.1);
\end{tikzpicture}\\
$\GTV$ &
$\frac{1}{2}\left( 2 + 3\sqrt{2} \right) \approx 3.121$&
$\frac{1}{2}\left( 1 + \sqrt{5} + 2\sqrt{2} \right) \approx 3.032$&
$\frac{1}{2}\left( 1 + \sqrt{13} + \sqrt{2} \right) \approx 3.010$\\
  \end{tabular}
  \caption{Illustration of geometric total variation on a simple black and white image. Even if the pixel values of the image are the same (represented by black and white disks), different triangulations of the pixels yield different $\GTV$. In this case, the right one is the optimal triangulation and is clearly more aligned with the underlying contour. Vectorizing or zooming the right triangulation will provide more interesting results. \label{fig-illustration-gtv}}
\end{table}

\RefAlgorithm{alg-optimize-gtv} is the main function that tries to
find a triangulation $T$ which is as close as possible to
$\GTV(s)$. It starts from any triangulation of $s$ (\RefLine{nl-init}
: we just split into two triangles each square between four adjacent
pixels). Then it proceeds by passes (\RefLine{nl-global-loop}). At
each pass (\RefLine{nl-local-loop}), every potential arc is checked
for a possible flip by a call to \textsc{CheckArc}
(\RefAlgorithm{alg-check-arc}) at \RefLine{nl-check-arc}. The arc is
always flipped if it decreases the $\GTV$ and {\em randomly} flipped
if it does not change the $\GTV$ (\RefLine{nl-flip-condition}). Arcs
close to the flip are queued (\RefLine{nl-flip-around}) since  their
  further flip may decrease the $\GTV$ later. The algorithm stops when
  the number of flips decreasing the $\GTV$ is zero (flips keeping the
  same $\GTV$ are not counted).

\begin{algorithm}
  \caption{Function \textsc{OptimizeGTV}  outputs a triangulation which is as close as possible to $\GTV(s)$. It builds a trivial triangulation of the pixels of $s$ then optimize its $\GTV$ by flipping its edges in a greedy and randomized way.\label{alg-optimize-gtv}}
  \fct{\textsc{OptimizeGTV}( $s$ : Image ) : Triangulation} \;
  \var{$T$ : Triangulation} \tcc*[h]{the triangulation that is optimized}\;
  \var{$Q$ : Queue$<$Arc$>$}\tcc*[h]{the queue of arcs currently being checked}\;
  \var{$Q'$ : Queue$<$Arc$>$}\tcc*[h]{the queue of arcs to be checked after}\;
  \Begin
    {
      $T \leftarrow \textsc{TrivialTriangulation}(\textsc{Pixels}(s))$\;
      \tcc{All arcs may be flipped at the beginning.}
      \nllabel{nl-init}
      \lFor{all arcs $a$ of $T$}{Push $a$ in $Q'$}
      \nllabel{nl-global-loop}
      \Repeat{$n = 0$}{
        \textsc{Swap}$(Q,Q')$\;
        $n \leftarrow 0$ \quad \tcc*[h]{Counts the number of flips that decrease $\GTV$}\;
        \nllabel{nl-local-loop}
        \While{$\neg \textsc{IsEmpty}(Q)$}{
          Pop $a$ from $Q$ \;
         $c \leftarrow \textsc{CheckArc}(s, T, a )$ \;        \nllabel{nl-check-arc}          
          \If{$c > 0 \text{~or~} (c = 0 \text{~and~} \textsc{FlipCoin}() = \mathrm{Heads})$}{ \nllabel{nl-flip-condition}
            \lIf{$c > 0 $}{$ n \leftarrow n+1$}
            Push arcs of two faces around $a$ in $Q'$\; \nllabel{nl-flip-around}
            \textsc{Flip}$(T,a)$ \quad \tcc*[h]{After, $a$ represents the flipped arc}\;
          }
        }
      }
      \ret{T}\;
    }  
\end{algorithm}

Function \textsc{CheckArc} (\RefAlgorithm{alg-check-arc}) checks if
flipping some arc would decrease the $\GTV$. It first checks if the
arc/edge is flippable at \RefLine{nl-is-flippable} (e.g. not on the
boundary). Then it is enough to check only one arc per edge
(\RefLine{nl-one-arc-per-edge}). Afterwards the edge may be flipped
only if the four points surrounding the faces touching the edge form a
strictly convex quadrilateron (\RefLine{nl-is-convex}). Finally it
computes the local $\GTV$ of the two possible decompositions of this
quadrilateron using formula \Eq{eq-gtv-expression} and outputs the
corresponding case (from  \RefLine{nl-compute-gtv}).

\begin{algorithm}
  \caption{Function \textsc{CheckArc} checks if flipping the arc $a$ of triangulation $T$ decreases $\GTV(T,s)$ (returns 1), does not change $\GTV(T,s)$ (returns 0), or returns $-1$ in all the other cases. \label{alg-check-arc}}
  \fct{\textsc{CheckArc}( $s$ : Image, $T$ : Triangulation, $a$ : Arc ) : Integer} \;
  \Begin
    {
      \lIf{$\neg \textsc{isFlippable}(T,a)$}{\ret{-1}} \nllabel{nl-is-flippable}
      $P \leftarrow \textsc{VerticesOfFacesAroundArc}(T,a)$ \;
      \tcc{$P[0]$ is Tail$(T,a)$, $P[2]$ is Head$(T,a)$, $P[0]P[1]P[2]$ and $P[0]P[2]P[3]$ are the two faces having $a$ in common.}
      \lIf{$P[0] < P[2]$}{\ret{-1}} \nllabel{nl-one-arc-per-edge}
      \lIf{$\neg \textsc{isConvex}(P)$}{\ret{-1}} \nllabel{nl-is-convex}
      $E_\mathrm{cur} = \NormK{ \DGrad s (P[0]P[1]P[2]) } + \NormK{ \DGrad s (P[0]P[2]P[3]) }$ \;       \nllabel{nl-compute-gtv}
      $E_\mathrm{flip} = \NormK{ \DGrad s (P[0]P[1]P[3]) } + \NormK{ \DGrad s (P[1]P[2]P[3]) }$ \;
      \lIf{$E_\mathrm{flip} < E_\mathrm{cur}$}{\ret{1}}
      \lElseIf{$E_\mathrm{flip} = E_\mathrm{cur}$}{\ret{0}}
      \lElse{\ret{-1}}
    }
\end{algorithm}

Note that the {\em randomization} of flips in the case where the
energy $\GTV$ is not changed is rather important in practice. As said
above, it is not a ``convex'' energy since it is easy to find
instances where there are several optima (of course, our search space being
combinatorial, convexity is not a well defined notion). Randomization
helps in quitting local minima and this simple trick gives nice
results in practice.

\begin{figure}[t]
  \begin{tabular}{cccc}
    original & linear gradient \TV & linear gradient \GTV & crisp gradient \GTV \\
    \includegraphics[width=0.24\textwidth]{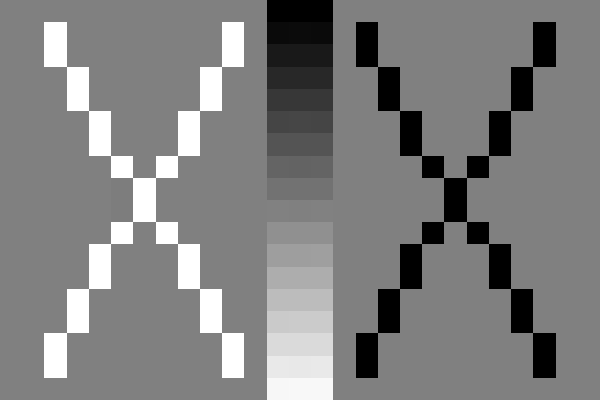}&
    \includegraphics[width=0.24\textwidth]{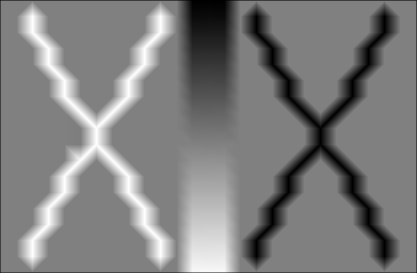}&
    \includegraphics[width=0.24\textwidth]{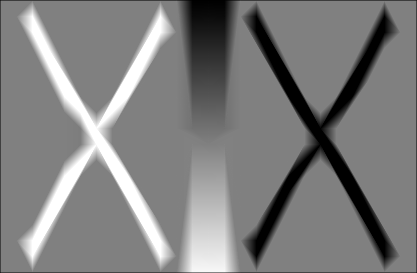}&
    \includegraphics[width=0.24\textwidth]{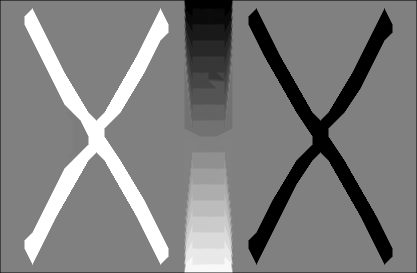}\\
    \includegraphics[width=0.24\textwidth]{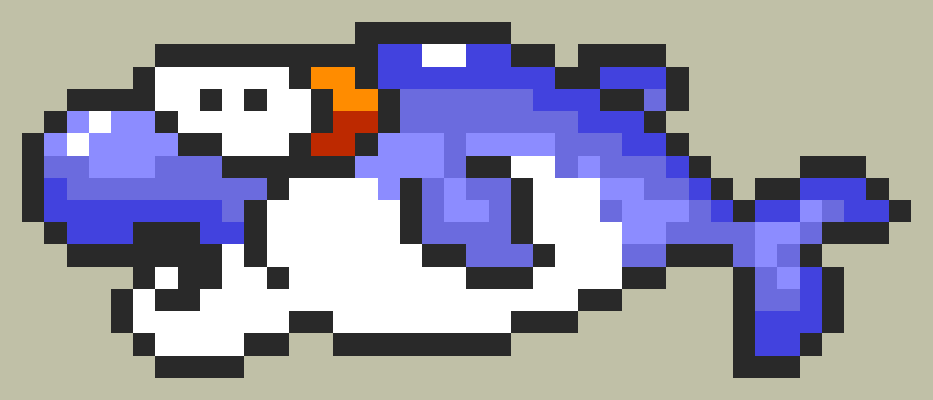}&
    \includegraphics[width=0.24\textwidth]{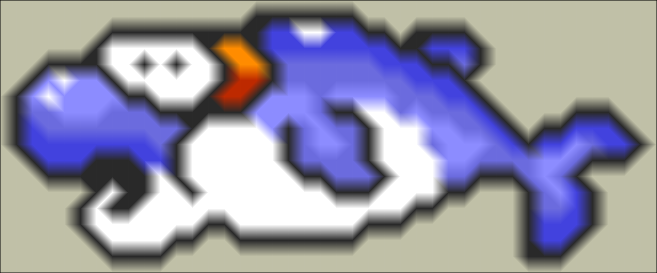}&
    \includegraphics[width=0.24\textwidth]{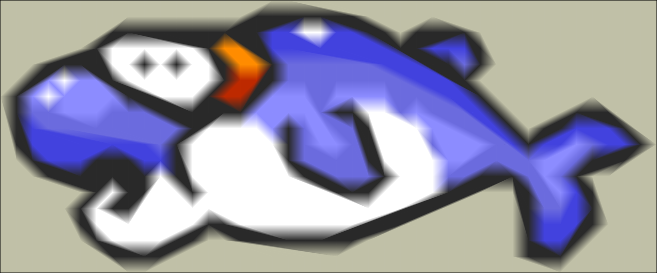}&
    \includegraphics[width=0.24\textwidth]{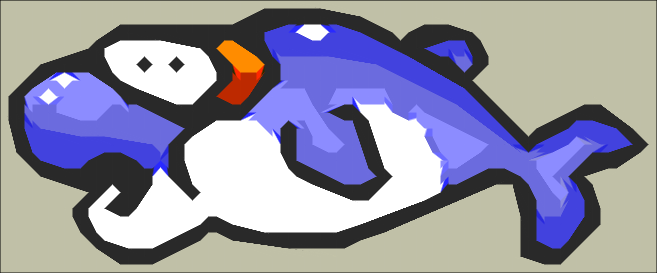}\\
    \includegraphics[width=0.24\textwidth]{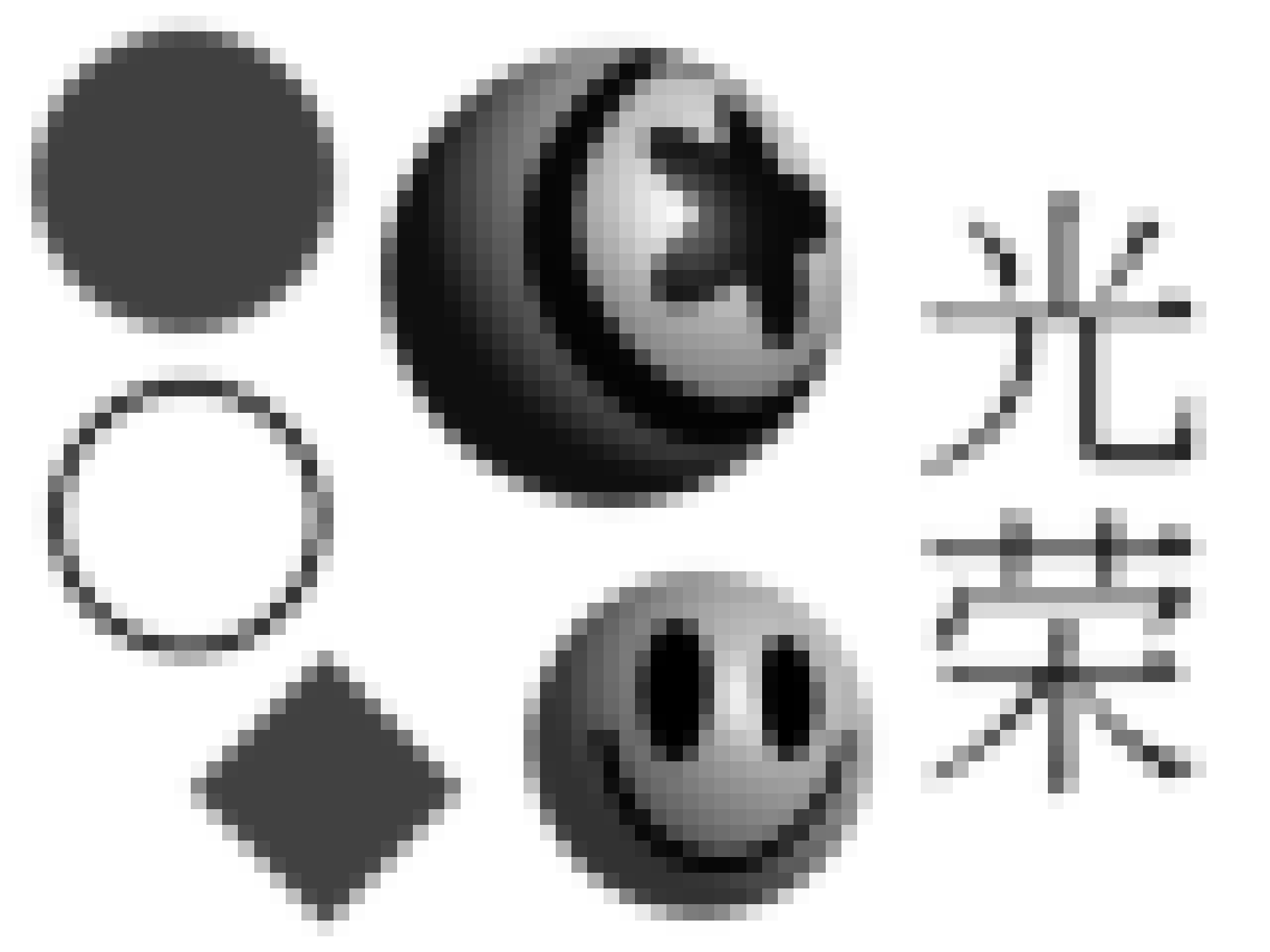}&
    \includegraphics[width=0.24\textwidth]{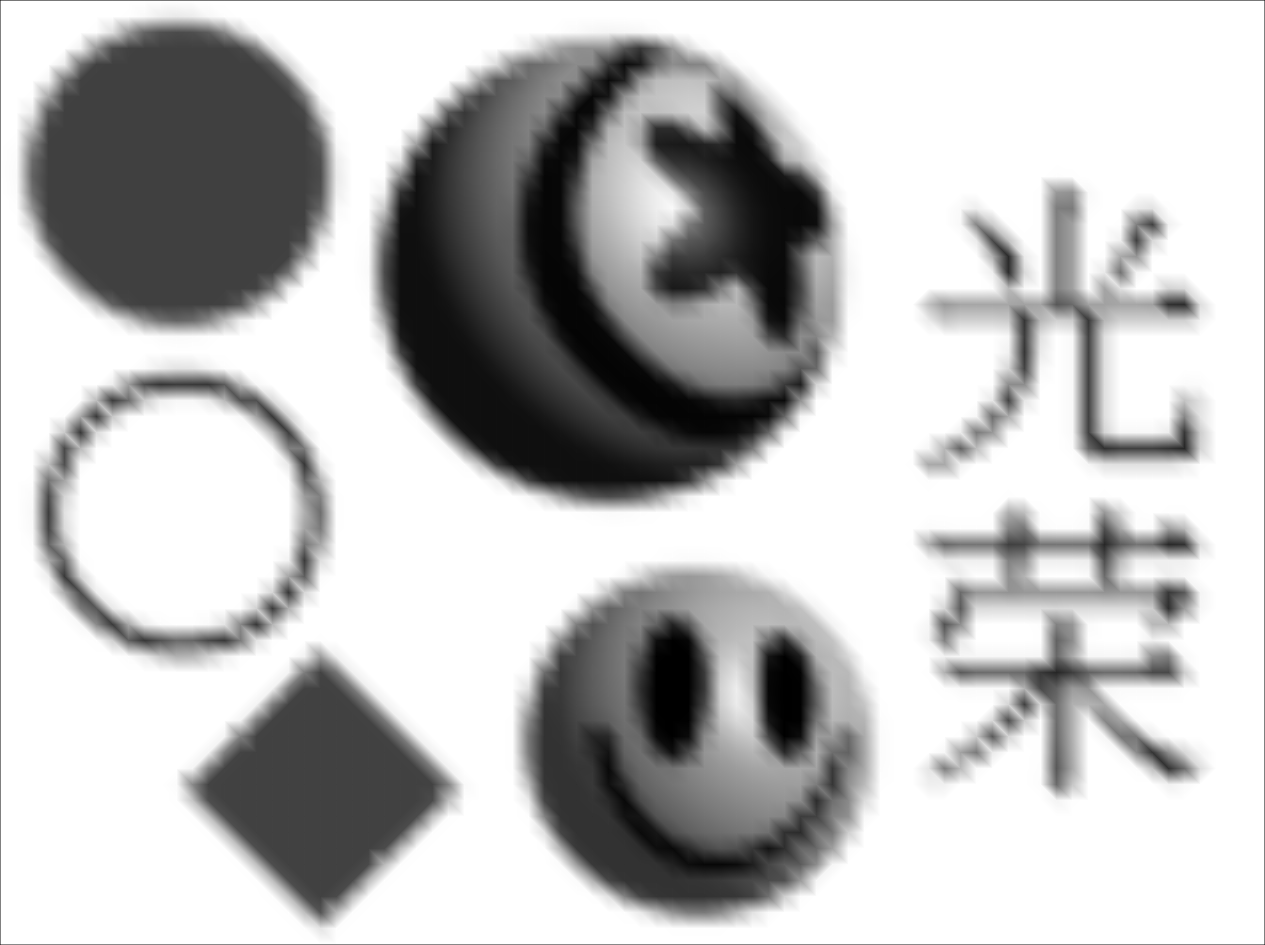}&
    \includegraphics[width=0.24\textwidth]{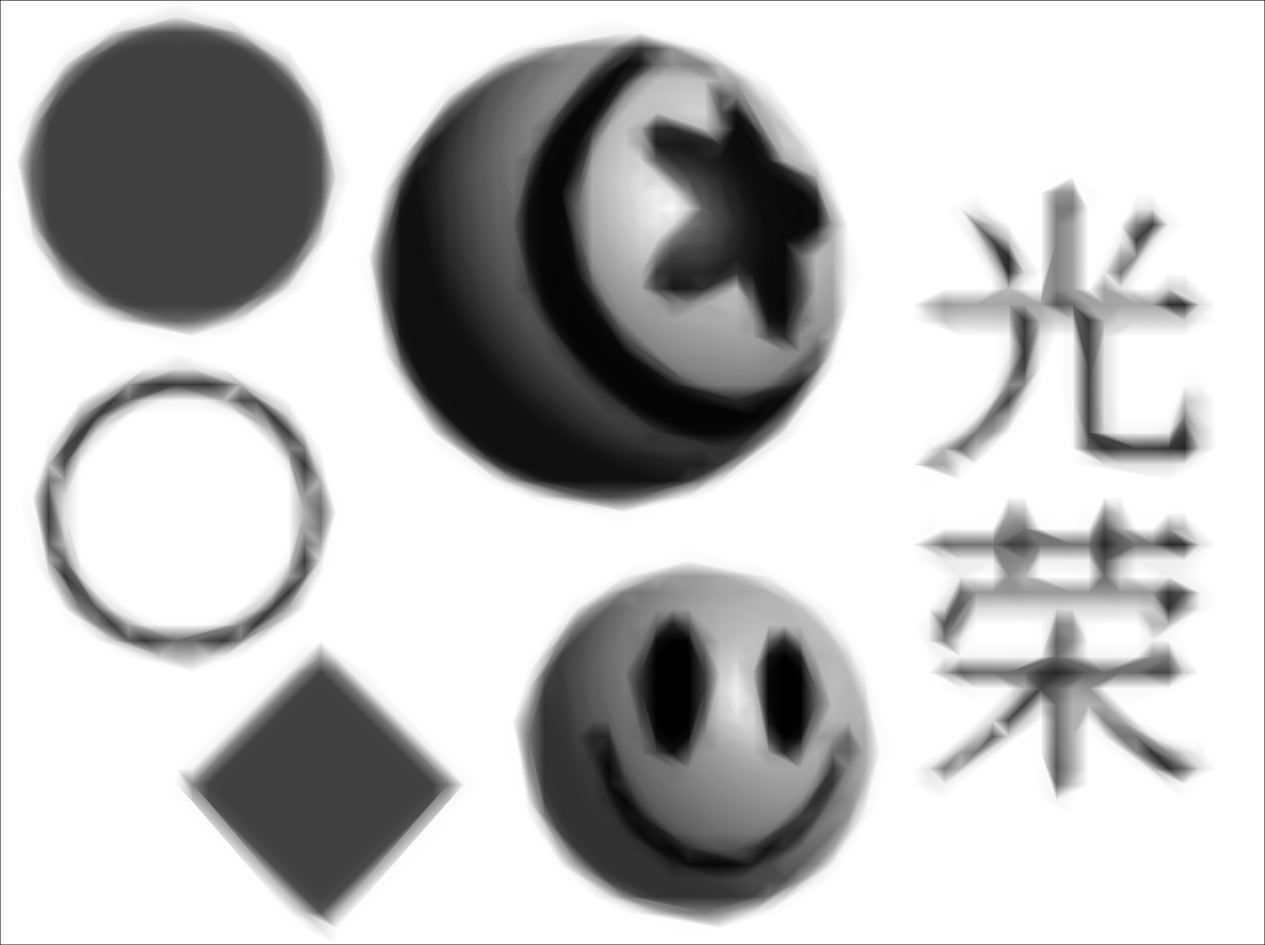}&
    \includegraphics[width=0.24\textwidth]{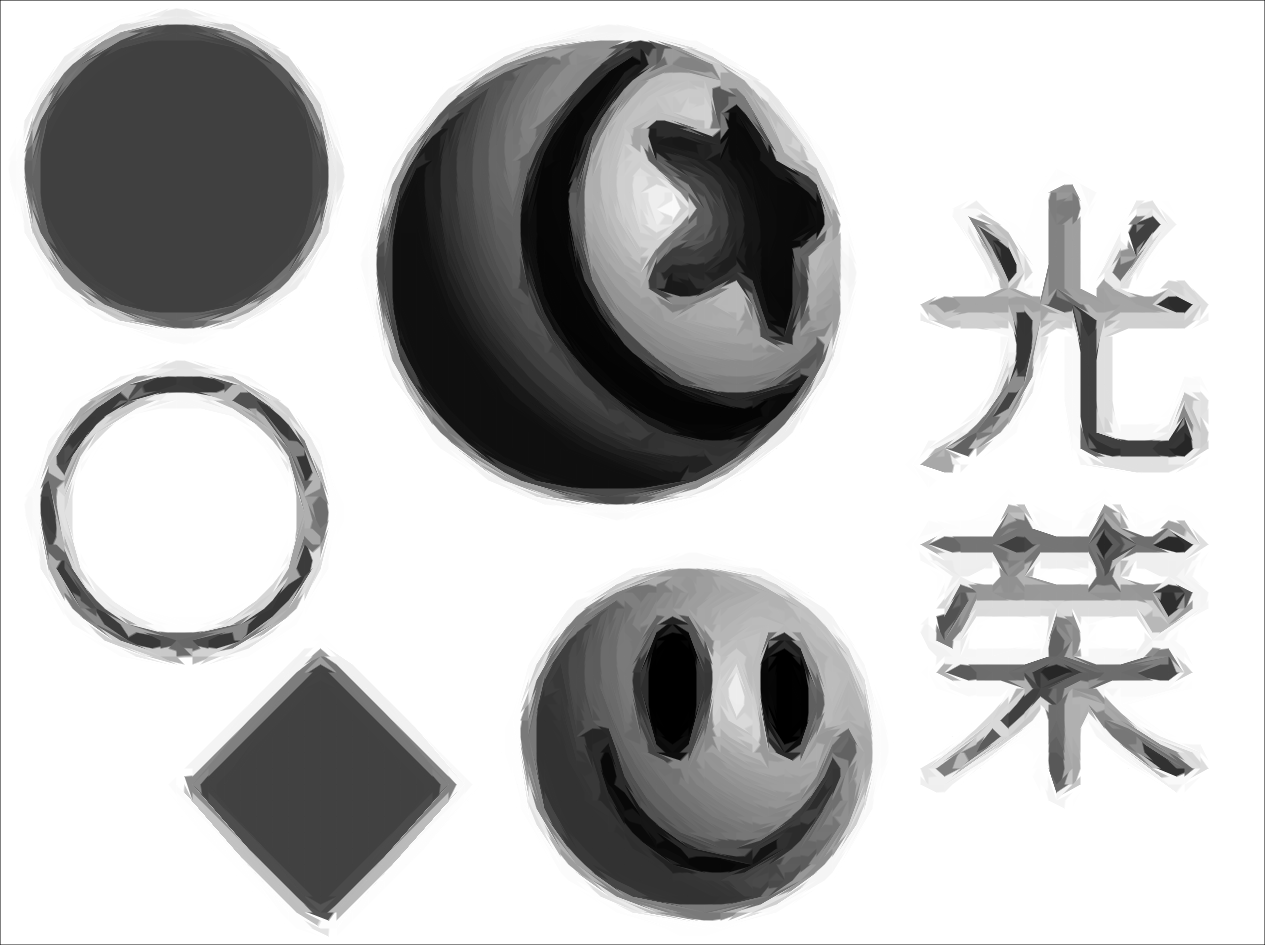}\\
    \includegraphics[width=0.24\textwidth]{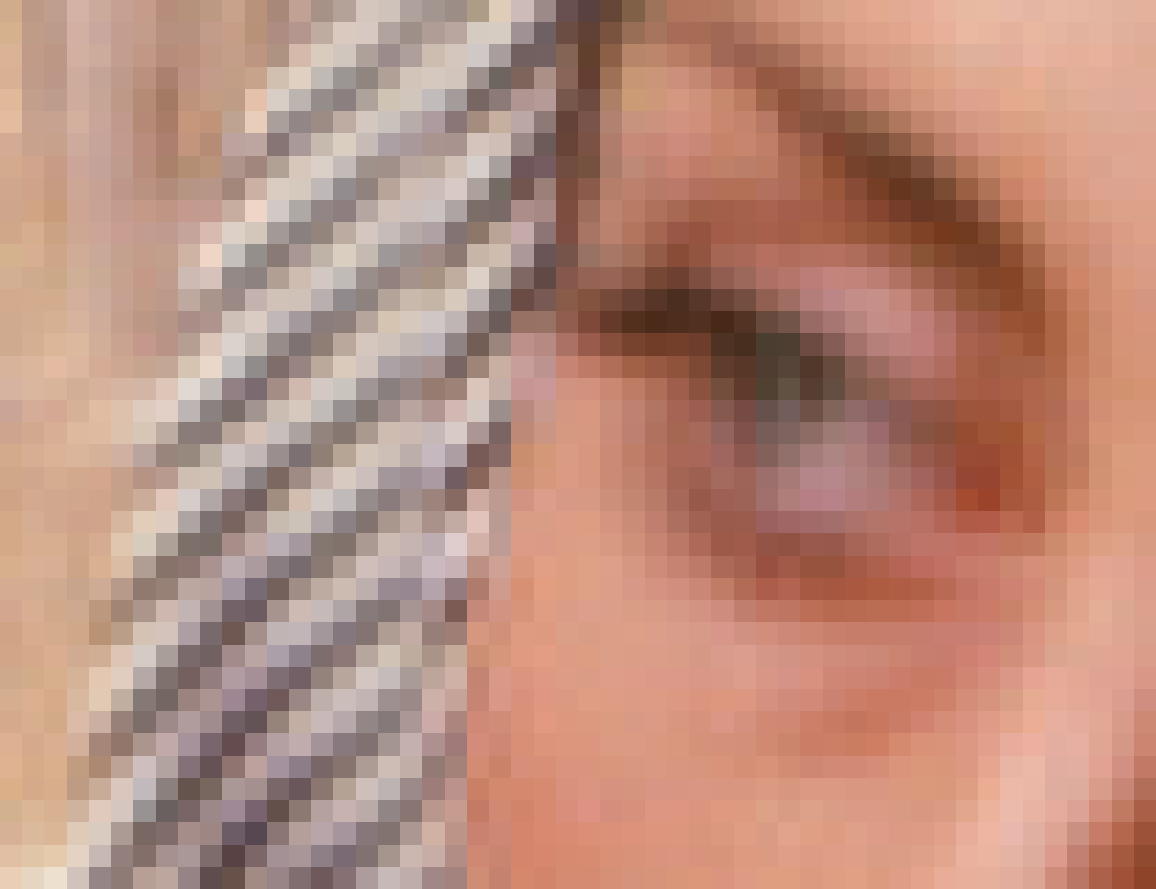}&
    \includegraphics[width=0.24\textwidth]{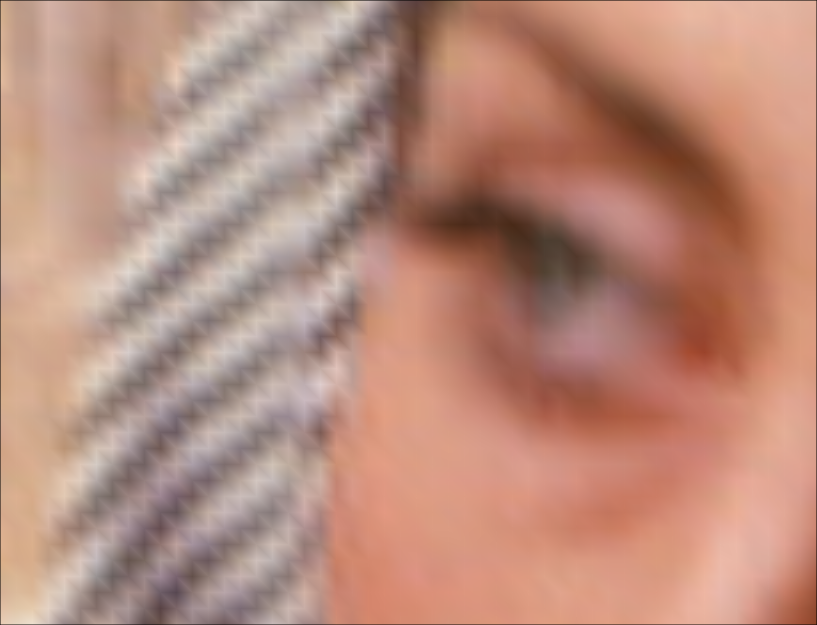}&
    \includegraphics[width=0.24\textwidth]{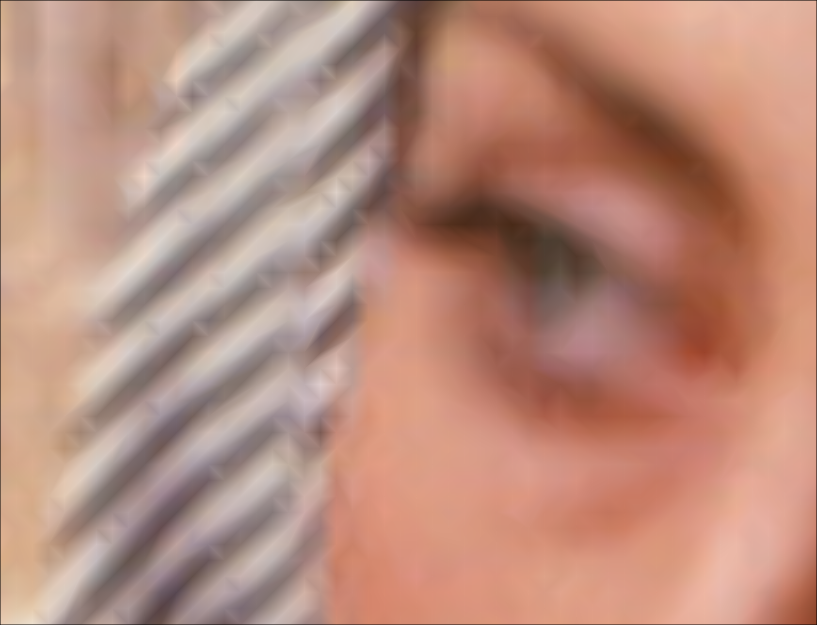}&
    \includegraphics[width=0.24\textwidth]{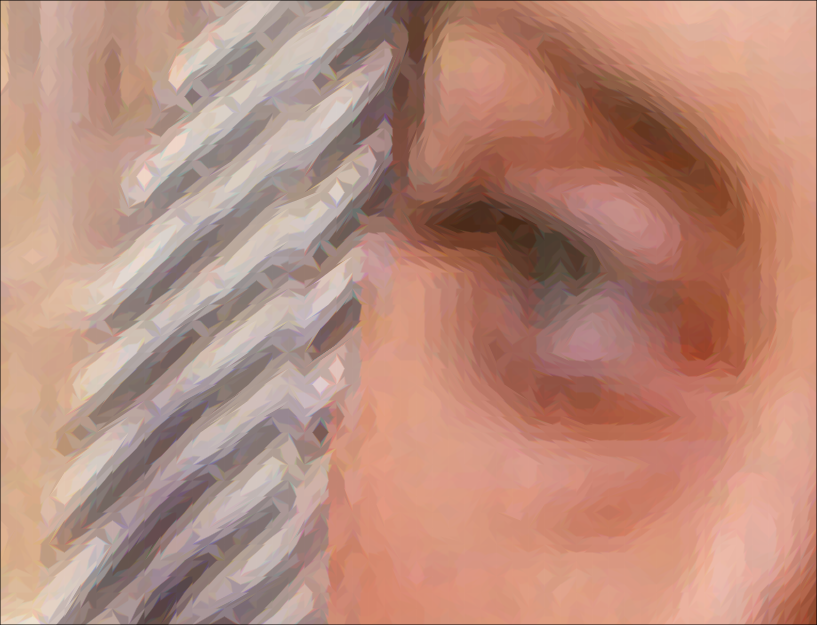}\\
    \includegraphics[width=0.24\textwidth]{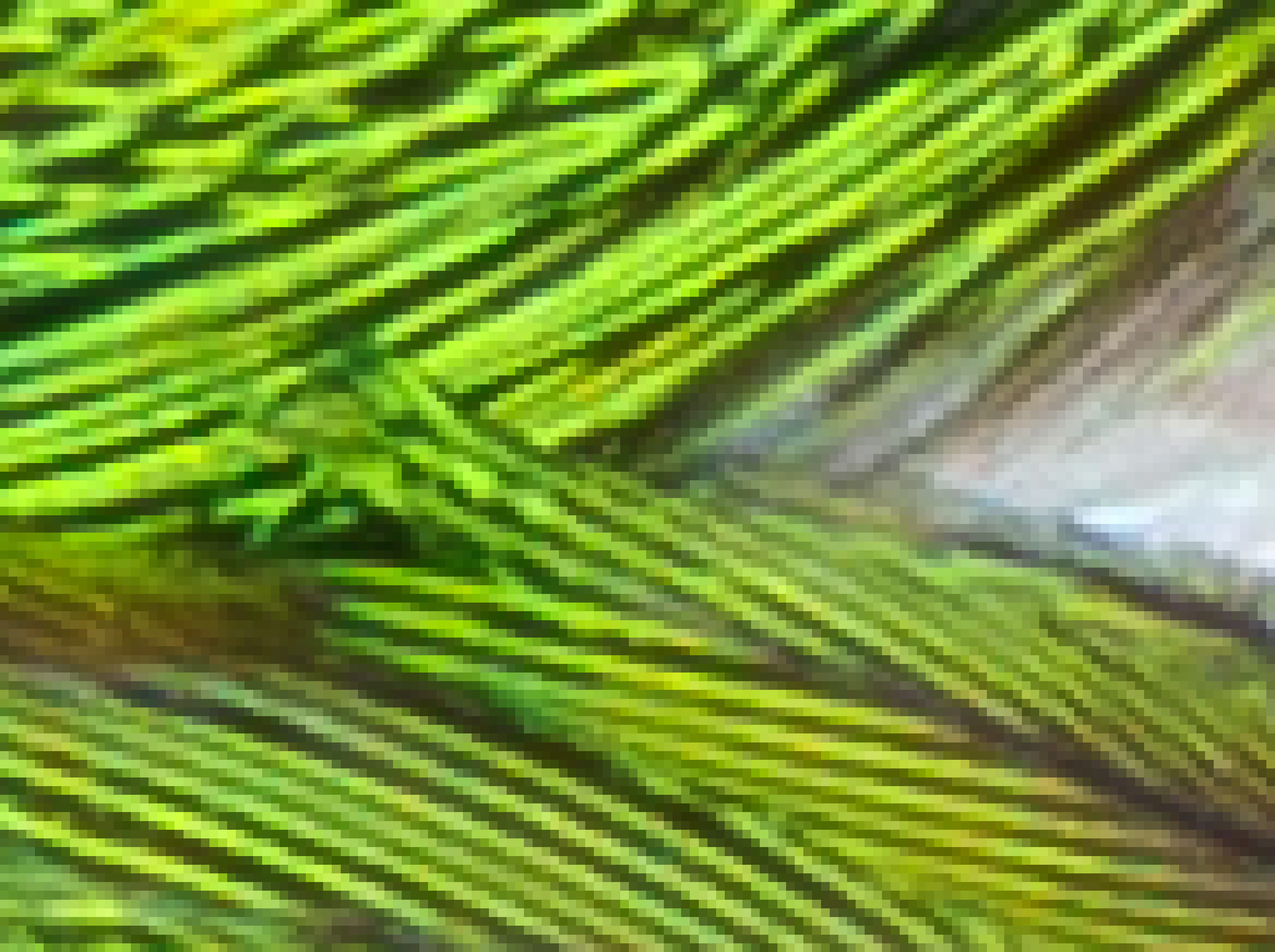}&
    \includegraphics[width=0.24\textwidth]{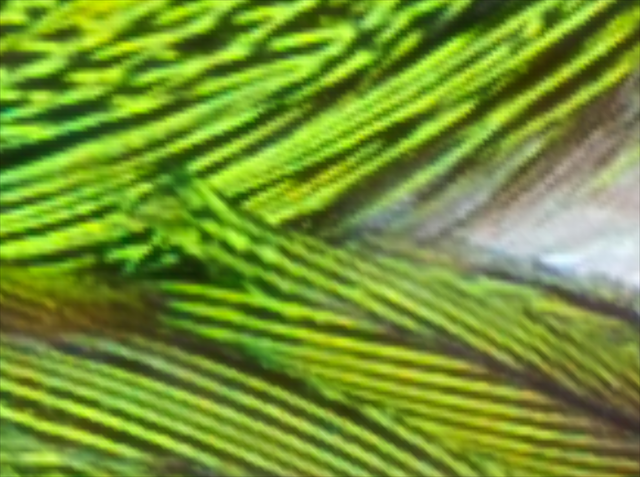}&
    \includegraphics[width=0.24\textwidth]{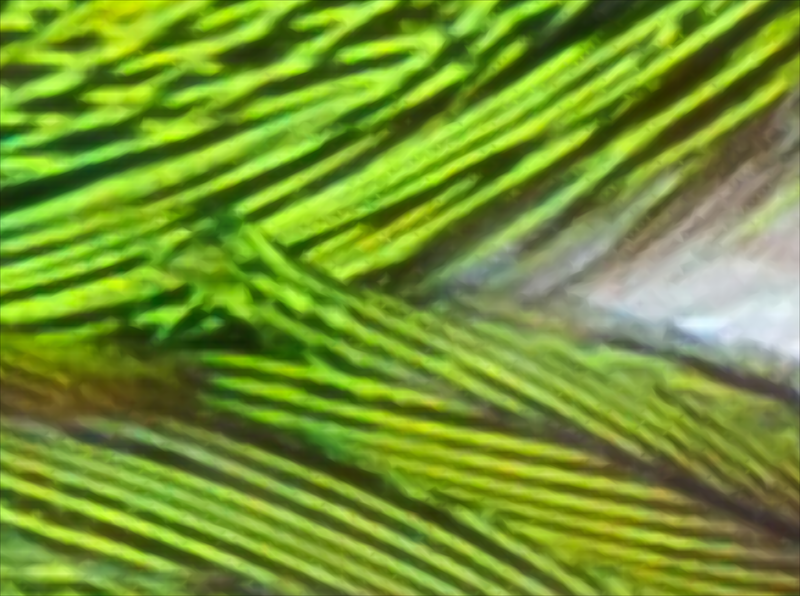}&
    \includegraphics[width=0.24\textwidth]{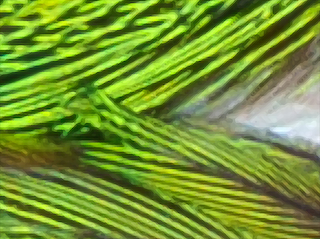}\\
  \end{tabular}
  \caption{\label{fig-gtv-zoom}Zoom $\times 16$ of the images in leftmost column: (middle left) displaying the trivial triangulation given by a simple discretization of $\TV$ painted with induced linear gradient over each triangle, (middle right) displaying the triangulation after $\GTV$ optimization painted with induced linear gradient over each triangle, (rightmost) same as before except that triangles are painted with a crisped version of the same gradient.}
\end{figure}

\RefFigure{fig-gtv-zoom} illustrates the capacity of geometric total
variation to capture the direction of image discontinuities. In this
simple application, we use the discrete gradient on each triangle to
display it with the corresponding linear shaded gradient. Hence we can
zoom arbitrarily any image. The results show that, if we stay with the
initial triangulation (as does standard discretization of $\TV$),
zoomed image are not great. On the contrary, results are considerably
improved if we use the optimized triangulation of $\GTV$. Last, we can
simply render triangles with a crisped version of this
gradient. Results are very nice in case of images with few colors like
in pixel art.

\section{Contour Regularization}
\label{sec-contour-reg}

\begin{wrapfigure}[6]{r}{0.28\textwidth}
\vspace{-0.07\textwidth}
  \includegraphics[width=0.23\textwidth]{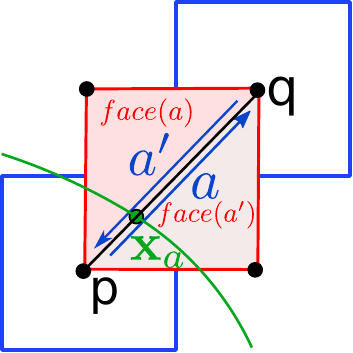}
\end{wrapfigure}
Contours within images are in-between pixels. In some sense they are
dual to the structure of the pixels. Since the Geometric Total
Variation has structured the relations between pixels, it is natural
to define the contours as a kind of dual graph to the optimal
triangulation $T^*$. We wish also that these contour lines align
nicely with discontinuities.

To do so, we introduce a variable $t_a$
on each arc whose value is kept in $\lbrack 0,1 \rbrack$. If the arc
$a$ is the oriented edge $(\vp,\vq)$ then the position of the contour
crossing this edge will be $\vx_a = t_a \vp + (1-t_a) \vq$ (see the upper floating figure).

We denote by $a'$ the arc opposite to $a$, i.e. $(\vq,\vp)$. The face
that is to the left of $a$ is denoted $\Face{a}$ while the one to its
right is $\FaceP{a}$.  We will guarantee that, at the end of each
iteration, $\vx_a = \vx_{a'}$. We associate to each arc $a=(\vp,\vq)$
a {\em dissimilarity weight} $w_a := \NormK{s(\vq) - s(\vp)} =
w_{a'}$.
We introduce also a point $\vb_f$ for each face $f$ of $T$, which will
lie at a kind of weighted barycenter of the vertices of $f$.

\RefAlgorithm{alg-reg-contour}  regularizes these points and provides a graph of
contours that is a meaningful vectorization of the input bitmap image
$s$. Note that the function \textsc{Intersection}$(a,\vy,\vz)$ returns
the parameter $t$ such that $\vx_a$ lies at the intersection of straight line
$( \vy \vz )$ and the arc $a$.

\begin{algorithm}
  \caption{\label{alg-reg-contour} Function \textsc{RegularizeContours}
    iteratively moves contour points and barycenters such that they align with the edges of the triangulation that delineate image discontinuities.}
  \fct{\textsc{RegularizeContours}( $s$ : Image, $T$ : Triangulation )} \;
  \KwResult{Positions of contour points on arcs and barycenter on faces}
  \Begin
    {
      \lFor{every face $f=\vp\vq\vr$ of $T$}{
        $\vb_f \leftarrow \frac{1}{3}(\vp+\vq+\vr)$}       \nllabel{nl-rg-init}
      \For{every arc $a$ of $T$}{ \nllabel{nl-rg-init2}
        \lIf{$\neg \textsc{isUpdateable}(a)$}{$t^{(0)}_a \leftarrow \frac{1}{2}$}
        \lElse{$t^{(0)}_a \leftarrow \textsc{Intersection}(a,\vb_{\Face{a}},\vb_{\FaceP{a}})$}
      }
      $n \leftarrow 0$\;
      \Repeat{$\max_{\text{arc~}a\text{~of~}T} | t^{(n)}_a - t^{(n-1)}_a | < \epsilon$}{       \nllabel{nl-rg-loop}
        \For(\tcc*[h]{Update barycenters}){every face $f$ of $T$}{         \nllabel{nl-rg-up-barycenters}
          $\vb_f \leftarrow \frac{1}{2}( \vb_f + \sum_{a \in \partial f} w_a \vx_a / \sum_{a \in \partial f} w_a )$\;
        }
        \For(\tcc*[h]{Update contours}){every arc $a$ of $T$ with $\textsc{isUpdateable}(a)$}{         \nllabel{nl-rg-up-contours}
          $t^{(n)}_a \leftarrow \frac{1}{2}( t^{(n)}_a + \textsc{Intersection}(a,\vb_{\Face{a}},\vb_{\FaceP{a}})$\;
        }
        \tcc{Average displacements along each edge according to area.}
        \For{every arc $a=\vp\vq$ of $T$ with $\vp < \vq$ and $\textsc{isUpdateable}(a)$}{  \nllabel{nl-rg-avg-contours}
            $(\alpha,\alpha') \leftarrow \frac{1}{2}(1+1/(6\textsc{Area}(a)),1+1/(6\textsc{Area}(a')))$\; \nllabel{nl-rg-area-formula}
            $t \leftarrow \frac{1}{2}(\alpha t^{(n)}_a + 1 - \alpha' t^{(n)}_{a'})$\;
            $(t^{(n+1)}_a,t^{(n+1)}_{a'}) \leftarrow (t,1-t)$ \;
        }
        $n \leftarrow n+1$\;
      }
      \ret{$(\vx,\vb)$}
    }
\end{algorithm}

First it starts with natural positions for $\vx$ and $\vb$
(resp. middle of arcs and barycenter of faces) at
\RefLine{nl-rg-init} and \RefLine{nl-rg-init2}. Then it proceeds iteratively until stabilization
in the loop at \RefLine{nl-rg-loop}. Each iteration consists of three
steps: (i) update barycenters such that they are a convex combination
of surrounding contour points weighted by their dissimilarities
(\RefLine{nl-rg-up-barycenters}), (ii) update contour points such that
they lie at the crossing of the arc and the nearby barycenters
(\RefLine{nl-rg-up-contours}), (iii) average the two displacements
along each edge according to respective area
(\RefLine{nl-rg-avg-contours}), thus guaranteeing that $\vx_a=\vx_{a'}$ for every arc.
The area $\textsc{Area}(a)$ associated to an arc $a$ is the area of
the quadrilateron formed by the tail of $a$, barycenters
$\vb_{\Face{a}}$ and $\vb_{\FaceP{a}}$ and $\vx_a$.  Note that the
coefficients $\alpha$ and $\alpha'$ computed at
\RefLine{nl-rg-area-formula} have
value 1 whenever the area is $\frac{1}{6}$.

\begin{figure}
  \begin{tabular}{ccc}
    & before regularization & after regularization\\
    \rotatebox{90}{\hspace{0.4cm}no $\GTV$} &
    \includegraphics[width=0.45\textwidth]{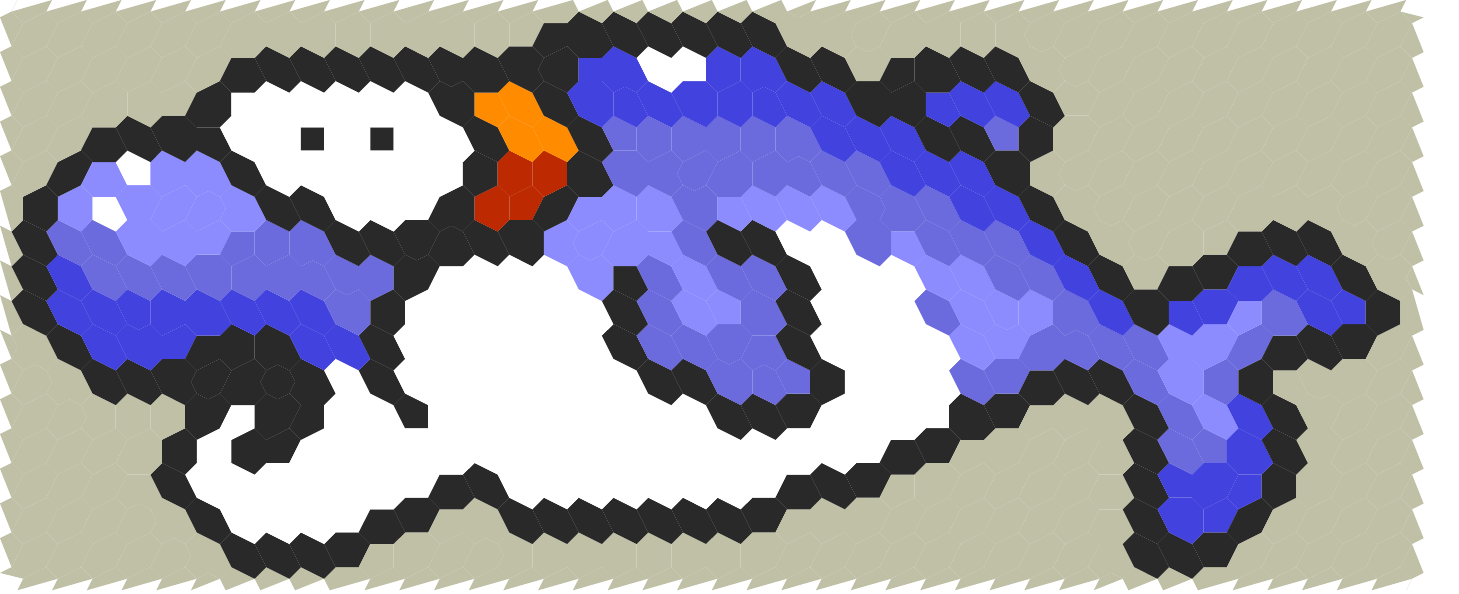}&
    \includegraphics[width=0.45\textwidth]{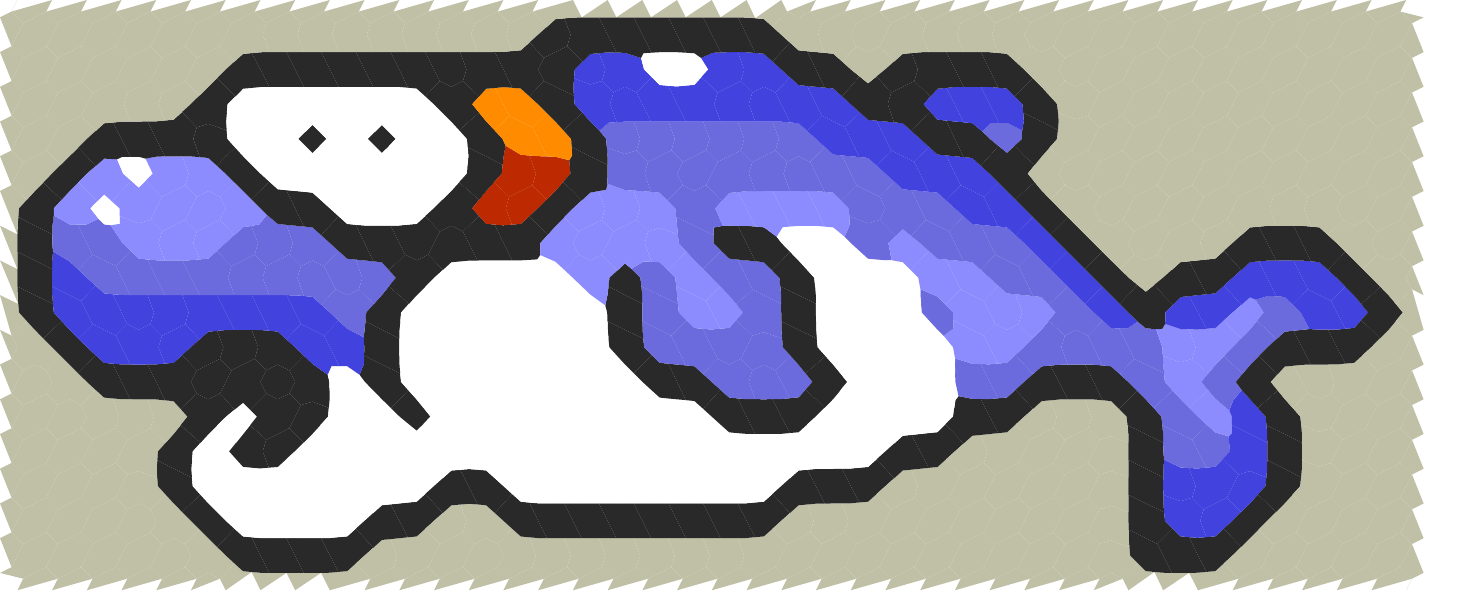}\\
    \rotatebox{90}{\hspace{0.4cm}with $\GTV$} &
    \includegraphics[width=0.45\textwidth]{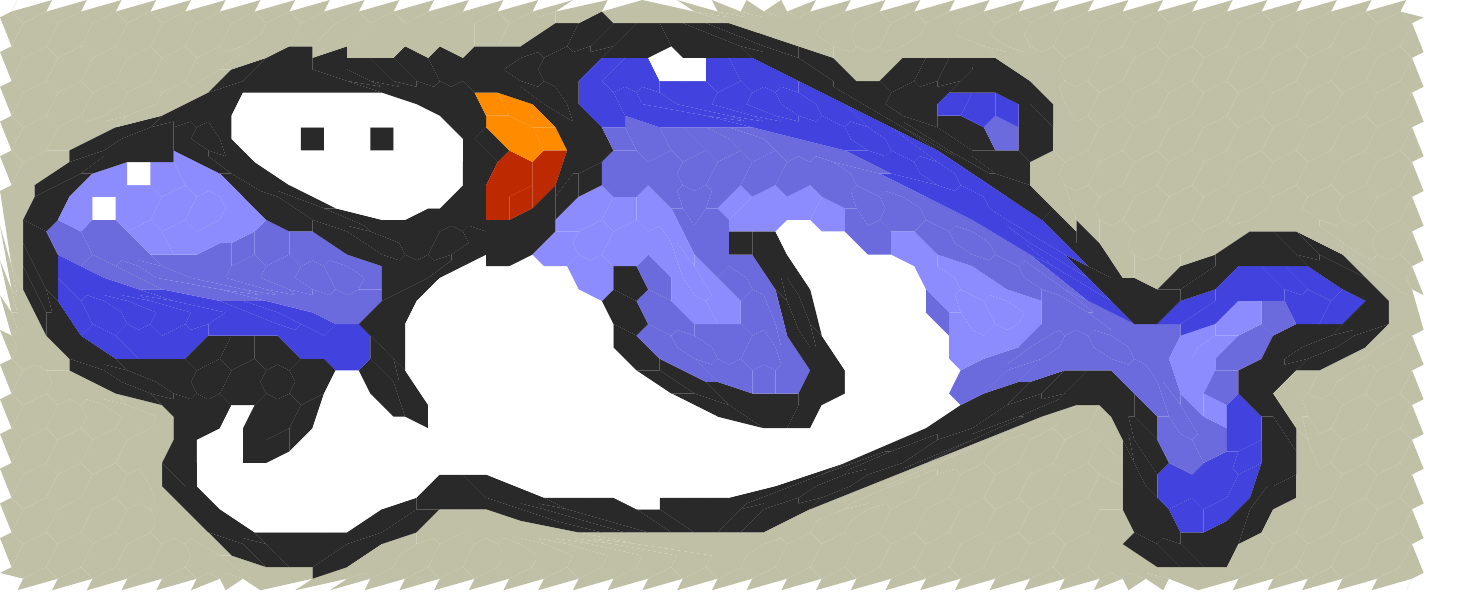}&
    \includegraphics[width=0.45\textwidth]{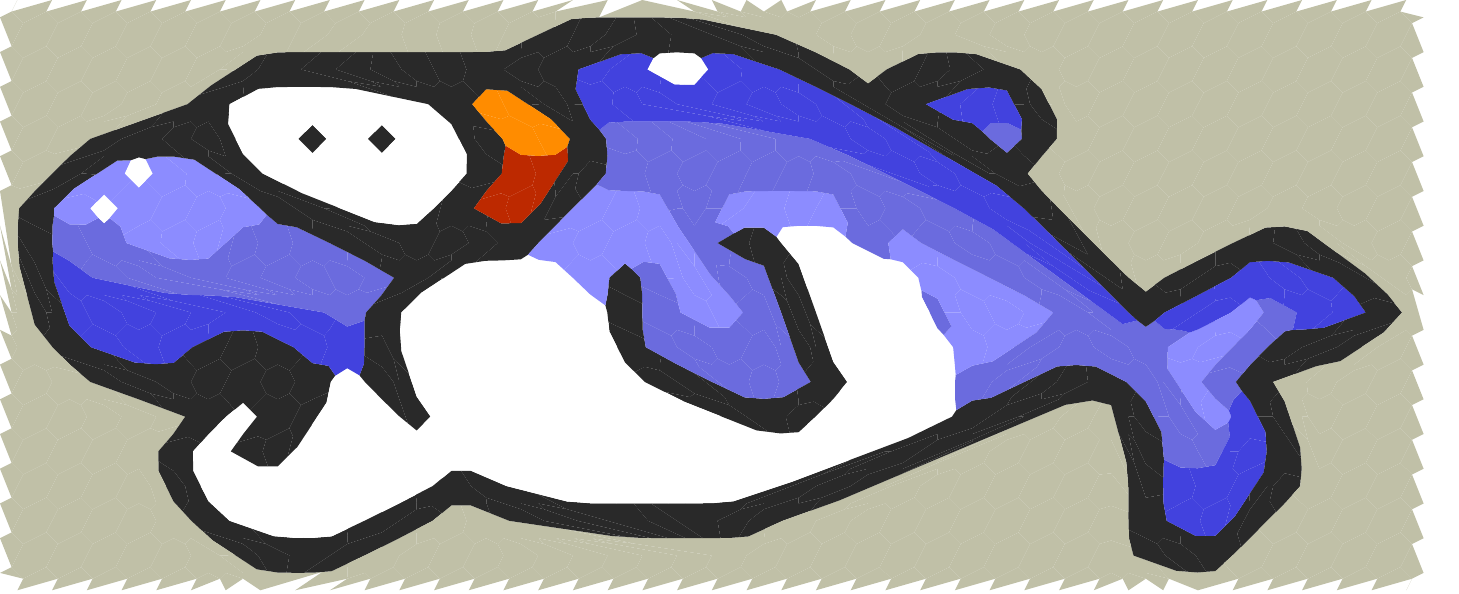}\\
  \end{tabular}
  \caption{\label{fig-dual-mesh} Displays contour mesh: left column before regularization, right column after regularization with \RefAlgorithm{alg-reg-contour}. Top row shows contour meshes when using the initial trivial triangulation. Bottom row shows contour meshes when using the triangulation $T^*$ that optimize $\GTV(s)$. Note that boundary triangles are displayed in white.}
\end{figure}

Since this process is always computing convex combinations of points
with convex constraints, it converges quickly to a stable point. In
all our experiments, we choose $\epsilon=0.001$ and the process
converges in a dozen of iterations. \RefFigure{fig-dual-mesh}
illustrates it. The {\em contour mesh} is defined simply as follows:
there is one cell per pixel, and for each pixel $\vp$, its cell is
obtained by connecting the sequence of points $\vx_{a_0},
\vb_{\Face{a_0}},\vx_{a_1}, \vb_{\Face{a_1}}\ldots$ for every arc
$a_0,a_1,\ldots$ coming out of $\vp$. Each cell of the contour mesh is
displayed painted with the color of its pixel. It is clear that the
contour mesh is much more meaningful after $\GTV$ optimization, and
its further regularization remove some artefacts induced by the
discretization. Last but not least, our approach guarantees that
sample points always keep their original color.

\section{Raster Image Zooming with Smooth Contours}
\label{sec-2nd-order}

Contour meshes as presented in \RefSection{sec-contour-reg} are easily
vectorized as polylines. It suffices to gather cells with same pixel
color as a regions and extract the common boundaries of these
regions. Furthermore, these polylines are easily converted to smooth
splines. We will not explore this track in this paper but rather
present a raster approach with similar objectives and features.

From the image $s$ with lattice domain $\Omega$, we wish to build a
{\em zoomed image} $s'$ with lattice domain $\Omega'$. If the zoom
factor is the integer $z$ and $\Omega$ has width $w$ and height $h$,
then $\Omega'$ has width $z(w-1)+1$ and height $z(h-1)+1$. The
canonical injection of $\Omega$ into $\Omega'$ is $\Injz: (x,y)
\mapsto (zx,zy)$. We use two auxiliary binary images $S$ (for
similarity image) and $D$ for (for discontinuity image) with domain
$\Omega'$. We also define the {\em tangent $\vT_a$} at an arc $a$ as
  the normalization of vector $\vb_{\Face{a}} - \vb_{\FaceP{a}}$.

The zoomed image $s'$ is constructed as follows:

\begin{description}
\item[Similarity set] We set to $1$ the pixels of $S$ that are in
  $\Injz(\Omega)$. Furthermore, for every arc $a=(\vp,\vq)$, if
  $w_a=0$ then the digital straight segment between $\Injz(\vp)$ and
  $\Injz(\vq)$ is also set to $1$. Last, we set the color $s'$ at
  these pixels to their color in $s$.
\item[Discontinuity set] For every face $f$, we count the number $n$
  of arcs whose weight is not null. If $n=0$ we simply set
  $D(\Injz(\vb_f))=1$. $n=1$ is impossible. If $n=2$, let $a_1$ and
  $a_2$ be the two arcs with dissimilarities. We set $D(\vp)=1$ for
  every pixel $\vp \in \Omega'$ that belongs to the digitized Bezier
  curve of order 3, that links the points $\Injz(\vx_{a_1})$ to
  $\Injz(\vx_{a_2})$, is tangent to $\vT_{a_1}$ and $\vT_{a_2}$, and
  pass through point $\Injz(\frac{1}{2}(\vb_f+I))$, with $I$ the
  intersection of the two lines defined by $\vx_{a_i}$ and tangent
  $\vT_{a_i}$, $i=1,2$. If $n=3$, for every arc $a$ of $f$, we set
  $D(\vp)=1$ for every pixel $\vp \in \Omega'$ that belongs to the
  digitized Bezier curve of order 2 connecting $\Injz(\vx_{a})$ to
  $\Injz(\vb_f)$ and tangent to $\vT_{a}$. Last we set the color $s'$
  at all these pixels by linear interpolation of the colors of pixels
  surrounding $s$.
\item[Voronoi maps] We then compute the voronoi map $\Vor{S}$
  (resp. $\Vor{D}$), which associates to each pixel of $\Omega'$ the
  closest point $\vp$ of $\Omega'$ such that $S(\vp)=1$ (resp. such
  that $D(\vp)=1$).
\item[Image interpolation] For all pixel $\vp$ of $\Omega'$ such that
  $S(\vp)=0$ and $D(\vp)=0$, let $\vq=\Vor{S}(\vp)$ and
  $\vq'=\Vor{D}(\vp)$. We compute the distances $d = \| \vq - \vp \|$
  and $d' = \| \vq' - \vp \|$. We use an amplification factor $\beta
  \in [0,1]$: when close to $0$, it tends to make linear shaded
  gradient while close to $1$, it makes contours very crisp.
  \begin{align*}
    s'(\vp) = \left\{
    \begin{array}{l}
      (1-\frac{2d'}{d+d'})s'(\vq') + \frac{2d'}{d+d'}( \beta s'(\vq) +(1-\beta)s'(\vq') ) \text{~when~}d' \le d, \\
      (1-\frac{2d}{d+d'})s'(\vq) + \frac{2d}{d+d'}( \beta s'(\vq) +(1-\beta)s'(\vq') ) \text{~otherwise}.
    \end{array}
    \right.
  \end{align*}
  In experiments, we always set $\beta=0.75$ which gives good results,
  especially for image with low quantization. Of course, many other
  functions can be designed and the crispness could also be
  parameterized locally, for instance as a function of the $\GTV$ of
  the triangle.
\item[Antialiasing discontinuities] To get images that are slightly
  more pleasant to the eye, we perform a last pass where we antialias
  every pixel $\vp$ such that $D(\vp)=1$. We simply perform a weighted
  average within a $3 \times 3$ window, with pixels in $S$ having
  weight $4$, pixels in $D$ having weight $0.25$ and other pixels
  having weight 1.
\end{description}

\begin{figure}[t]
  \begin{tabular}{cc}
    \includegraphics[width=0.48\textwidth]{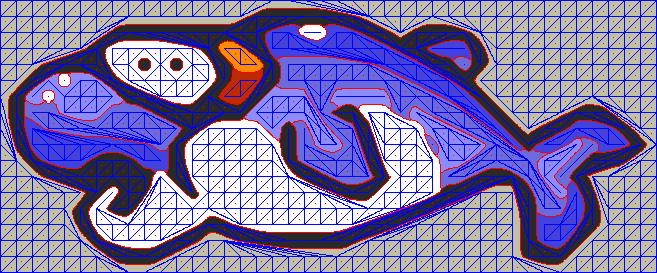}&
    \includegraphics[width=0.48\textwidth]{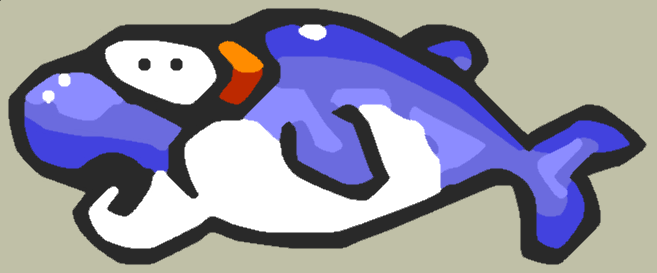}\\
  \end{tabular}
  \caption{\label{fig-smooth-raster-zooming}Illustration of raster image zooming with smooth contours. On the left, similarity set is drawn in blue while discontinuity set is drawn in red. The final result is displayed on the right.}
\end{figure}

\RefFigure{fig-smooth-raster-zooming} illustrates this method of
raster image zooming which provides crisp discontinuities that follow
Bezier curves. Comparing with \RefFigure{fig-dual-mesh}, image
contours are no more polygonal lines but look like smooth curves. Last
this method still guarantees that original pixel colors are kept.

\section{Experimental Evaluation and Discussion}
\label{sec-experiments}

\begin{figure}[h]
  \begin{tabular}{@{}ccccc}
  &   (a) original &  (b) Geom. stencil \cite{getreuer_image_2011bis}&(c) CNN \cite{shi_real-time_2016}  & (d) Our approach \\
  \rotatebox{90}{\hspace{0.25cm}x-shape}  &  \includegraphics[width=0.24\textwidth]{x-shapes-2-x16-bitmap.png}&
    \includegraphics[width=0.24\textwidth]{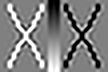}&
    \includegraphics[width=0.24\textwidth]{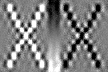}&
    \includegraphics[width=0.24\textwidth]{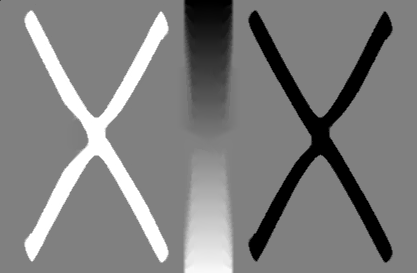}\\[-1mm]
    & & \scriptsize total time: $8\,ms $ & \scriptsize total time: $413\,ms $& \scriptsize total time: $218\,ms $ \\
    \rotatebox{90}{\hspace{0.01cm} dolphin} & \includegraphics[width=0.24\textwidth]{dolphin-x16-bitmap.png}&
    \includegraphics[width=0.24\textwidth]{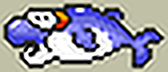}&
    \includegraphics[width=0.24\textwidth]{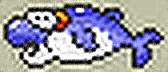}&
    \includegraphics[width=0.24\textwidth]{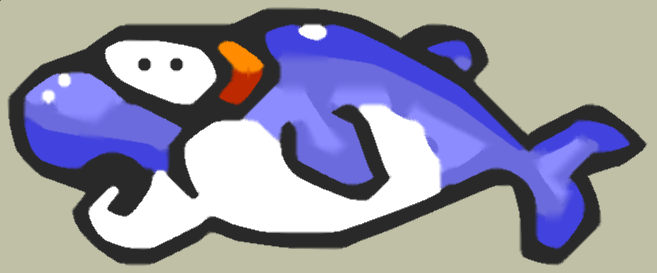}\\[-1mm]
    & &\scriptsize total time: $ 11\,ms $ & \scriptsize total time: $398\,ms $ & \scriptsize total time: $320\,ms $ \\
    \rotatebox{90}{ipol coarsened} & \includegraphics[width=0.24\textwidth]{ipol-coarsened-x16-bitmap.png}&
    \includegraphics[width=0.24\textwidth]{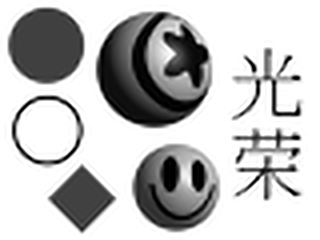}&
    \includegraphics[width=0.24\textwidth]{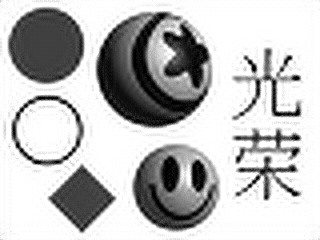}&
       \includegraphics[width=0.24\textwidth]{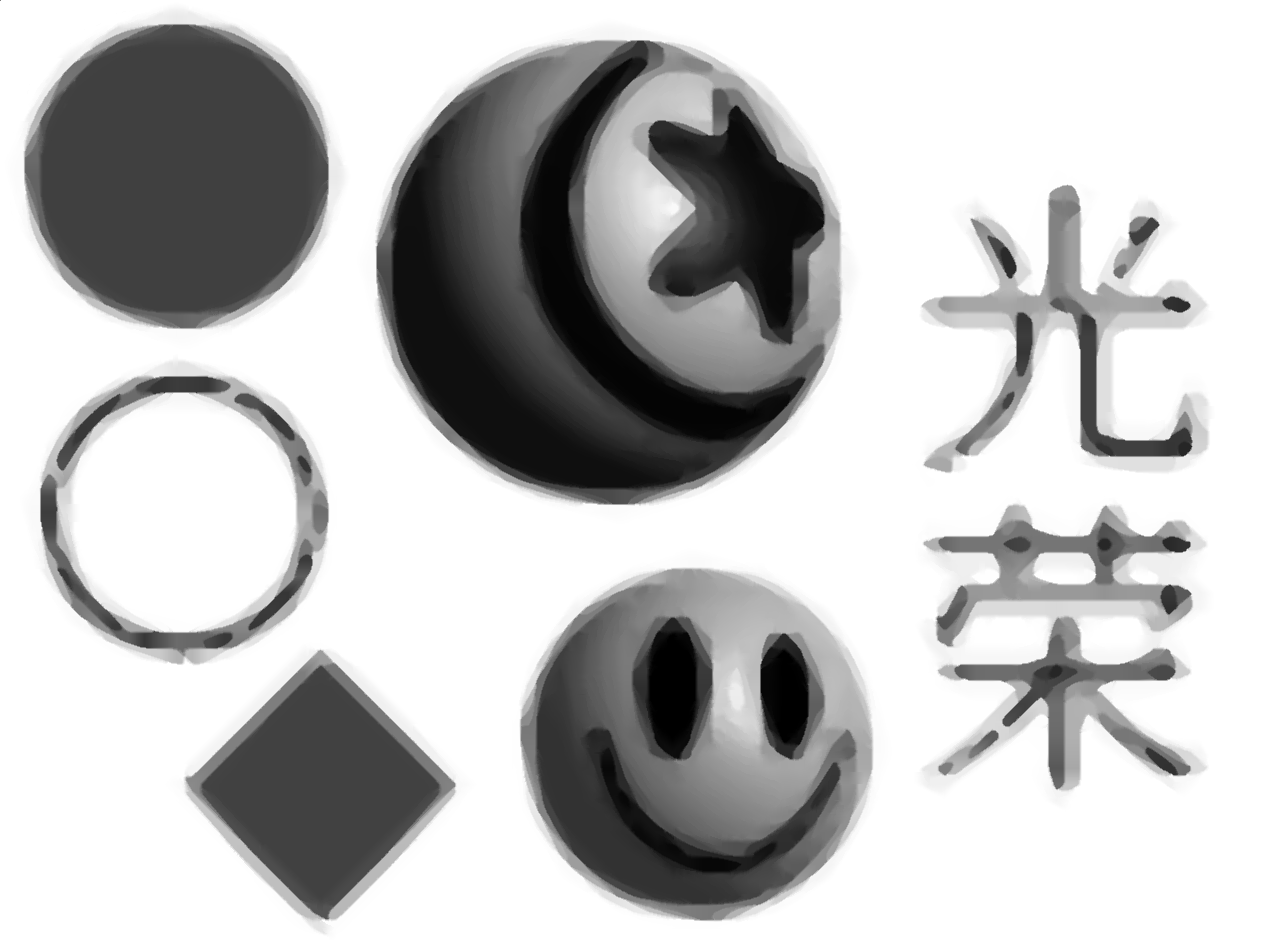}\\[-1mm]
    & &\scriptsize total time: $ 63\,ms $ & \scriptsize total time:$425\,ms $& \scriptsize total time: 2$032\,ms $ \\
    \rotatebox{90}{\hspace{0.25cm}barbara}& \includegraphics[width=0.24\textwidth]{barbara-eyes-x16-bitmap.png}&
    \includegraphics[width=0.24\textwidth]{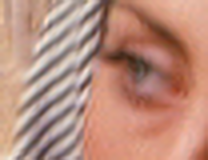}&
\includegraphics[width=0.24\textwidth]{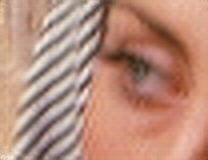}&
    \includegraphics[width=0.24\textwidth]{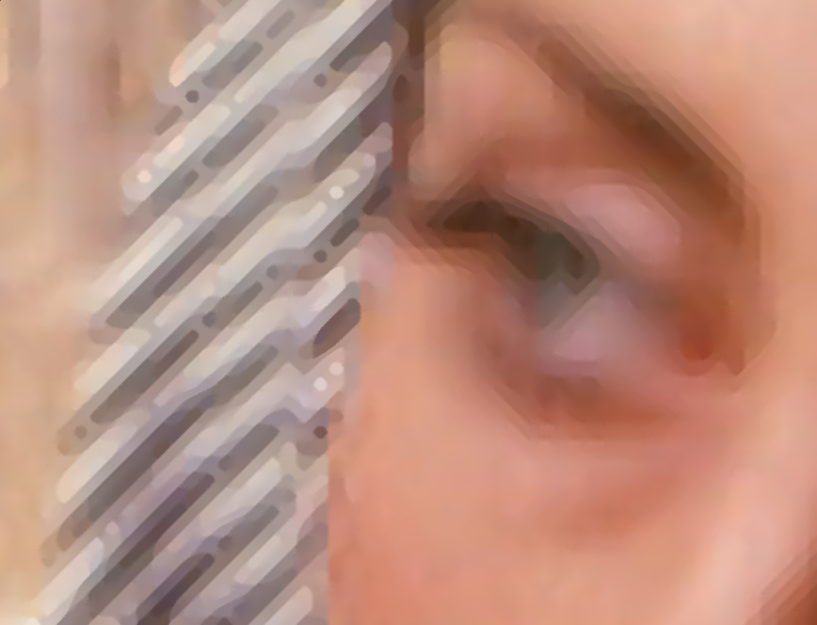}\\[-1mm]
    & & \scriptsize total time: $ 28\,ms $& \scriptsize total time: $412\,ms $ & \scriptsize total time: 1$029\,ms $ \\
    \rotatebox{90}{\hspace{0.5cm} ara}  &  \includegraphics[width=0.24\textwidth]{ara-detail-x16-bitmap.png}&    
    \includegraphics[width=0.24\textwidth]{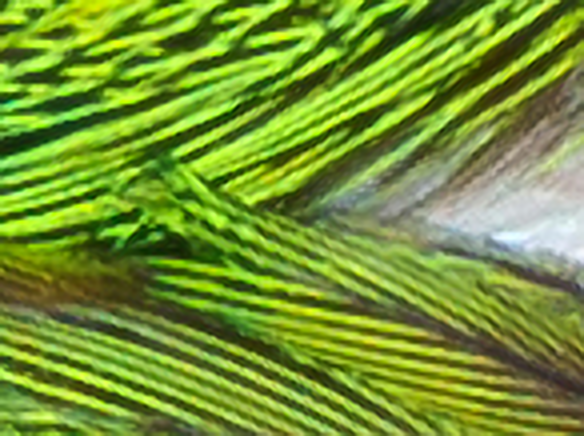}&
\includegraphics[width=0.24\textwidth]{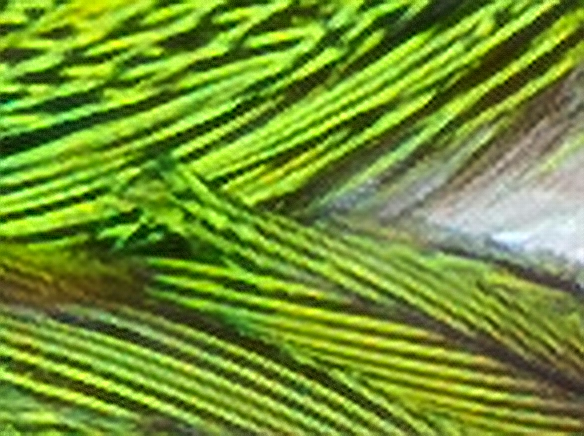}&
    \includegraphics[width=0.24\textwidth]{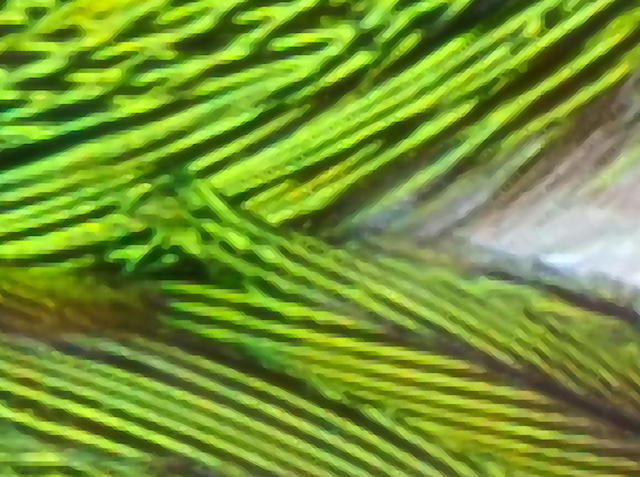}\\[-1.8mm]
    & &\scriptsize total time: $202\,ms$ &\scriptsize total time: $551\,ms$ & \scriptsize total time: $8524\,ms$ \\[-0.13cm]
    \multicolumn{5}{c}{\scriptsize The time measures were obtained on a 2.9 GHz \textit{Intel Core i7} for (c,d) and on IPOL server for (b).}
  \end{tabular}
  \caption{\label{fig-expe-comp1} Comparison of the proposed approach (d) with other methods on Geometric Stencils \cite{getreuer_image_2011bis} (b) and  based on Convolution Neural Network \cite{shi_real-time_2016} (c). }
\end{figure}

Our method can both produce vectorized image or make rasterized zoomed
images, either from camera pictures or tiny image with low
quantization like pixel art images. To measure its performance,
several other super-resolution methods have been tested on the
set of images given in the first column of 
\RefFigure{fig-expe-comp1}, with parameters kept constant across all
input images. First, we experimented the method based on geometric
stencils proposed by Getreuer \cite{getreuer_contour_2011} with default
parameters, which is implemented in the online demonstrator
\cite{getreuer_image_2011bis}. As shown in the second column of the
figure, results appear noisy with oscillations near edges, which are
well visible on pixel art images (e.g. x-shape or dolphin
images). Such defaults are also visible on ara or barbara image near
the white area. Another super-resolution method was experimented which
uses on a convolutional neural network \cite{shi_real-time_2016}. Like
the previous method, numerous perturbations are also visible both on
pixel art images (x-shape or dolphin) but also in
homogeneous areas close to strong gradients of ipol-coarsened image.
We have tried different parameters but they lead to images with
comparable quality. On the contrary, due to its formulation, our
method does not produce false contours or false colors, and works
indifferently on pixel art images or camera pictures.

\begin{figure}[h]
          \begin{tabular}{cc}
               Depixelizing  \cite{kopf_depixelizing_2011} & vector magic \cite{vectorMagic10} \\
               \includegraphics[width=0.2\textwidth]{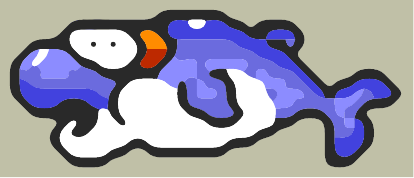}  &
              \includegraphics[width=0.2\textwidth]{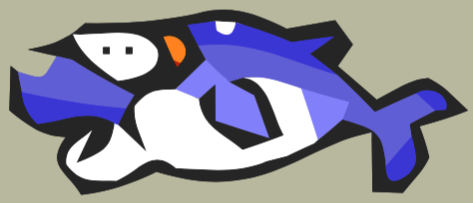} \\[-1mm]
              \scriptsize approx. time: $< 0.5\,s$ & \scriptsize approx. time: $2\,s$ \\
               \includegraphics[width=0.2\textwidth]{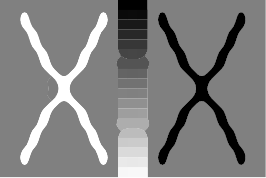}  & 
               \includegraphics[width=0.2\textwidth]{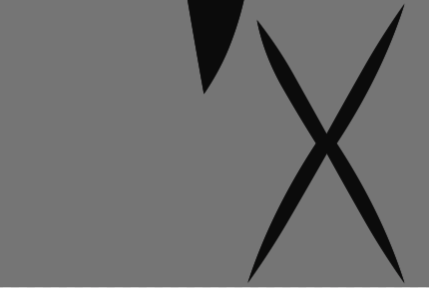} \\[-1mm]
              \scriptsize approx. time: $<0.5\,s$ & \scriptsize approx. time: $2\,s$ \\               
           \end{tabular}
           \begin{tabular}{cc}
           Roussos-Maragos \cite{getreuer_roussos-maragos_2011} & Potrace \cite{potrace}\\
           \includegraphics[width=0.275\textwidth]{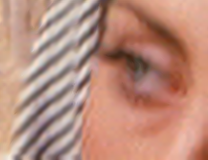}&
           \includegraphics[width=0.275\textwidth]{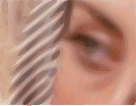}\\[-1mm]
            \scriptsize total time : $503\,ms$ & \scriptsize approx. time: $5\,s$
\end{tabular}
           \begin{tabular}{cccc}
               $Hq4x$ \cite{Stepin2003} & $Hq4x(Hq4x)$  & $Hq4x$ \cite{Stepin2003} & $Hq4x(Hq4x)$\\
             \includegraphics[height=0.175\textwidth]{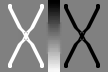}&
             \includegraphics[height=0.175\textwidth]{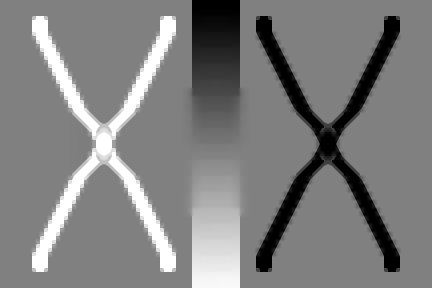}&
             \includegraphics[height=0.175\textwidth]{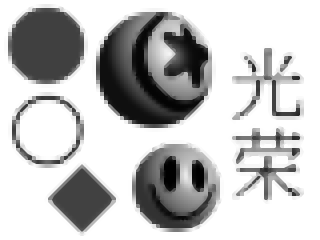}&
             \includegraphics[height=0.175\textwidth]{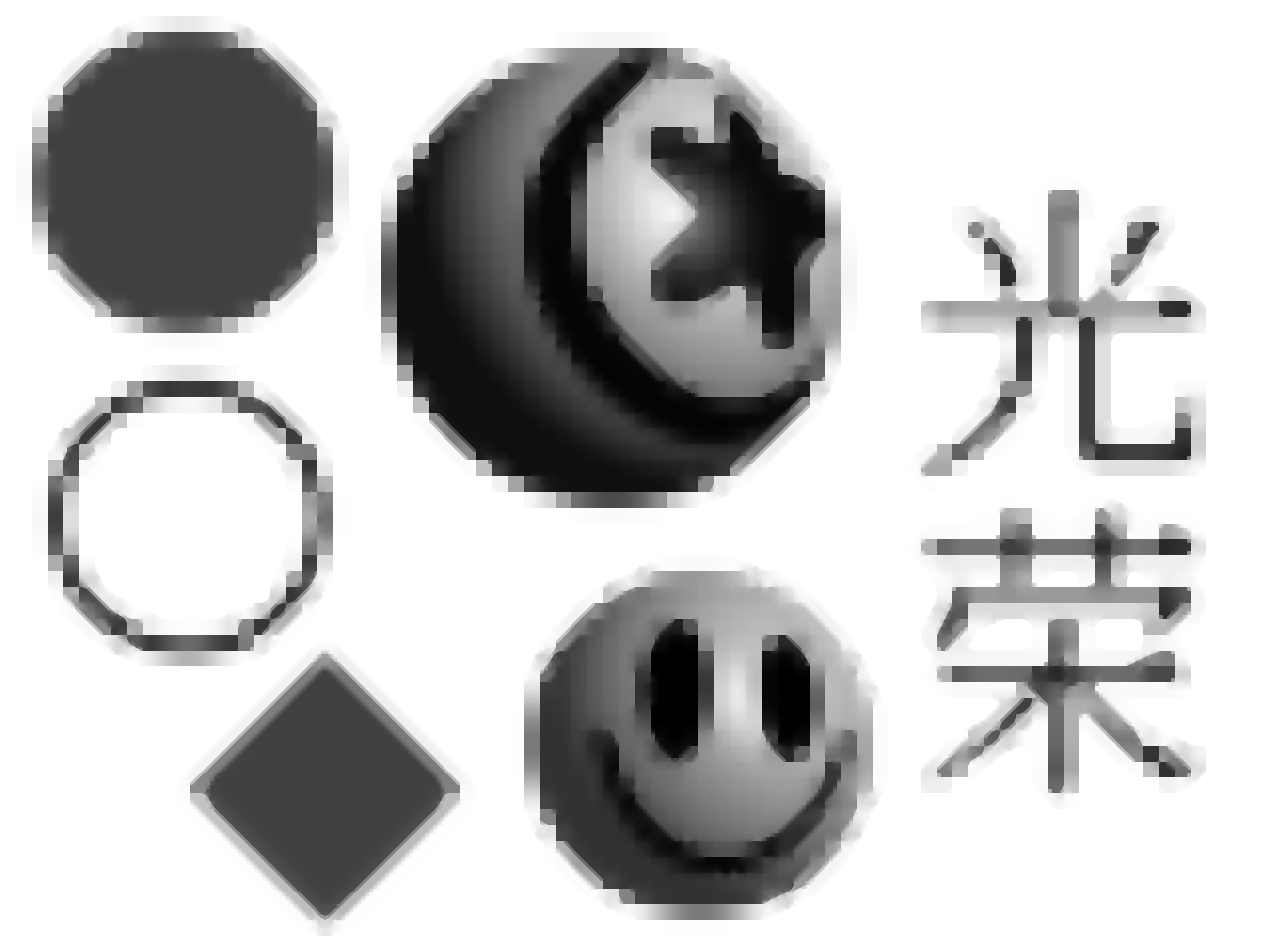}\\[-1mm]
             \scriptsize total time : $158\,ms$ & \scriptsize total time: $152\,ms$& \scriptsize total time : $142\,ms$  & \scriptsize total time: $320\,ms$
           \end{tabular}
   \caption{\label{fig-expe-comp2} Other complementary comparisons on five other approaches. The time measures were obtained on a 2.9 GHz \textit{Intel Core i7} for all experiments expect Roussos-Maragos obtained on IPOL server. We use the default parameters expect for Potrace: we select 512 passes with color option. }
\end{figure}

Other comparisons are presented on \RefFigure{fig-expe-comp2} in order
to give an overview of the behavior of five other methods. The two
first methods complement the comparisons on pixel art image with
respectively the depixelizing method proposed by Kopf and Lischinski
\cite{kopf_depixelizing_2011} implemented in \textit{Inkscape}, and a
commercial sofware proposed by \textit{Vector Magic Inc}
\cite{vectorMagic10}. Our method captures better the direction of
discontinuities of the underlying shape with less contour oscillations
(see the border of dolphin or x-shape). Two other methods were
tested on barbara image: Roussos and Maragos tensor-driven method
\cite{getreuer_roussos-maragos_2011} and Potrace vectorization
software \cite{potrace}. Again, results appear with oscillations
around strong gradients with Roussos-Maragos algorithm, while Potrace
software tends to smooth too much the image. Finally we also applied
the Hqx magnification filter proposed by Stepin \cite{Stepin2003} that
provides interesting zoom results but presents some artefacts (see for
instance the X center of Hq4x) and limited scale factor (i.e. 4). Note
that other comparisons can easily be done with the following online
demonstrator: {\small\url{https://ipolcore.ipol.im/demo/clientApp/demo.html?id=280}} 
and source code is available on a  \textit{GitHub} repository:
{\small\url{https://github.com/kerautret/GTVimageVect}}.

\section{Conclusion}

We have presented an original approach to the problem of image
vectorization and image super-resolution. It is based on a
combinatorial variational model, which introduces geometry into the
classical total variation model. We have compared our method both to
state-of-the-art vectorization and super-resolution methods and it
behaves well both for camera pictures and pixel art images. We have
also provided an online demonstrator, which allows users to reproduce
results or test our method with new images. In future works, we plan
to compare quantitatively our method with state-of-the-art techniques,
and also use our approach to train a CNN for zooming into pixel art
images.

\bibliographystyle{plain}
\bibliography{paper.bib}

\end{document}